\documentclass[reprint,showpacs,preprintnumbers,amsmath,amssymb,twocolumn,superscriptaddress]{revtex4}

	\usepackage{graphicx,dcolumn,bm}

\begin{document}
\title{Delayed Buckling and Guided Folding of Inhomogeneous Capsules}

    \author{Sujit S. Datta}
    \affiliation{These authors contributed equally to this work.}
    \affiliation{Department of Physics, Harvard University, Cambridge MA 02138, USA}           

    \author{Shin-Hyun Kim}
        \affiliation{These authors contributed equally to this work.}
            \affiliation{Harvard School of Engineering and Applied Sciences, Cambridge MA 02138, USA}
  \affiliation{Department of Chemical and Biomolecular Engineering, KAIST, Daejeon, South Korea}

    \author{Jayson Paulose}
        \affiliation{These authors contributed equally to this work.}
    \affiliation{Harvard School of Engineering and Applied Sciences, Cambridge MA 02138, USA}

    \author{Alireza Abbaspourrad}
    \affiliation{Harvard School of Engineering and Applied Sciences, Cambridge MA 02138, USA}
    \author{David R. Nelson}
    \affiliation{Department of Physics, Harvard University, Cambridge MA 02138, USA}
    \author{David A. Weitz}
    \affiliation{Department of Physics, Harvard University, Cambridge MA 02138, USA}
    \affiliation{Harvard School of Engineering and Applied Sciences, Cambridge MA 02138, USA}

\date{\today}

\begin{abstract}   
Colloidal capsules can sustain an external osmotic pressure; however, for a sufficiently large pressure, they will ultimately buckle. This process can be strongly influenced by structural inhomogeneities in the capsule shells. We explore how the time delay before the onset of buckling decreases as the shells are made more inhomogeneous; this behavior can be quantitatively understood by coupling shell theory with Darcy's law. In addition, we show that the shell inhomogeneity can dramatically change the folding pathway taken by a capsule after it buckles. 
\end{abstract}
\pacs{46.32.+x,46.70.De, 47.55.D-,47.56.+r,47.57.J-,62.20.mq,}
\maketitle

Many important natural or technological situations require understanding thin, spherical shells; examples include colloidal capsules for chemical encapsulation and release \cite{datta,mohwald,vanblaad}, biological cells \cite{vella,fletcher}, pollen grains \cite{eleni}, submersibles \cite{nash}, chemical storage tanks \cite{pederson}, nuclear containment shells \cite{pederson}, and even the earth's crust \cite{kearey}. In many cases, the utility of such a shell critically depends on its response to an externally-imposed pressure. For small pressures, a homogeneous, spherical shell, characterized by a uniform thickness, supports a compressive stress, and it shrinks isotropically. Above a threshold pressure, however, this shrinkage becomes energetically prohibitive; instead, the shell buckles, reducing its volume by forming a localized indentation at a random position on its surface. For the case of a homogeneous shell, this threshold pressure can be calculated using a linearized analysis of shell theory \cite{hutchinson,koiter}, while the exact morphology of the shell after it buckles requires a full nonlinear analysis \cite{gompper,knoche,vaziri}. However, many shells are inhomogeneous, characterized by spatially-varying thicknesses and elastic constants \cite{eleni,pinder,stupp,dufresne,julicher}. Such inhomogeneities can strongly influence how a shell buckles \cite{eleni,hutchinson,pauchard,pap,vernizzi, krenzke, bushnell}. Unfortunately, despite its common occurrence in real shells, exactly how inhomogeneity influences the onset of buckling, as well as the shell morphology after buckling, remains to be elucidated. A deeper understanding requires careful investigations of the buckling of spherical shells with tunable, well-defined, inhomogeneities. 

In this Letter, we use a combination of experiments, theory, and simulation to study the buckling of spherical colloidal capsules with inhomogeneous shells of non-uniform thicknesses. We show that the onset of buckling, above a threshold external osmotic pressure, is well described by shell theory; however, even above this threshold, the capsules do not buckle immediately. We find that the time delay before the onset of buckling decreases as the shells are made more inhomogeneous; these dynamics can be quantitatively understood by coupling shell theory with Darcy's law for flow through a porous capsule shell, even for highly inhomogeneous shells. Moreover, we find that the shell inhomogeneity guides the folding pathway taken by a capsule during and after buckling. We use these insights to controllably create novel colloidal structures using buckling. 

We fabricate monodisperse thin-shelled capsules using water-in-oil-in-water (W/O/W) droplets prepared by microfluidics \cite{shin, supp}. The inner and outer phases are a 10 wt \% solution of polyvinyl alcohol (PVA) of viscosity $\mu=13.5$ mPa-s, as measured using a strain-controlled rheometer, while the middle oil phase is a photo-polymerizable monomer solution. The PVA solution is less dense than the oil; as a result, after the droplets are produced and collected, the light inner water droplets gradually rise within them. This causes the oil to gradually thin on the top side of each droplet and thicken on the bottom \cite{bigal}. We exploit this effect to prepare capsules, with outer radius $R_{0}$, and spatially-varying shell thickness $h(\theta)\approx h_{0}-\delta\mbox{cos}\theta$; $\theta$ is measured from the top of the gravitationally-oriented shell, $h_{0}$ is the average shell thickness, and $\delta$ is the total distance moved by the inner droplet, as shown schematically in Fig. 1(a). The shell inhomogeneity can thus be quantified by the ratio $\delta/h_{0}$. We use UV light to polymerize the oil either as the capsules are produced {\it in situ}, or after different average waiting times, $t_{w}$ \cite{spread}; this enables us to prepare separate batches of capsules characterized by varying degrees of shell inhomogeneity \cite{finite,sem}. Some capsules are subsequently washed in de-ionized water. The shell is a solid characterized by a Young's modulus $E\approx600$ MPa \cite{jiang}; importantly, while this shell is impermeable to Na$^{+}$ and Cl$^{-}$ ions, it is permeable to water \cite{shin2}.

To probe their mechanical response, we subject inhomogeneous capsules, characterized by $t_{w}=1$ min, $\delta/h_{0}\approx0.2$ and $h_{0}/R_{0}=0.017$, to an external osmotic pressure by injecting and gently mixing $20\mu$L of the capsule suspension into a fixed volume of NaCl solution, $V_{NaCl}\approx130-400~\mu$L. We investigate the pressure dependence of buckling using NaCl concentrations in the range 0.063-2.165 M. Estimating the total volume of the injected capsules using optical microscopy allows us to calculate the final NaCl concentration of the outer phase, which then ranges from $c_{NaCl}=0.055-2.068$ M. These correspond to osmotic pressure differences across the shell of $\Pi = (2c_{NaCl} +\Pi_{out}-\Pi_{in}) \times N_{A}k_{B}T = 0.025-10.09$ MPa, where $N_{A}$ is Avogadro's constant, $k_{B}$ is Boltzmann's constant, $T\approx300$ K, and $\Pi_{out}$ and $\Pi_{in}$ are the measured osmolarities of the fluids outside and inside the capsules, respectively, in the absence of NaCl. For each batch of capsules studied, we monitor an average of 75 capsules over time using optical microscopy. 

\begin{figure}
\begin{center}
\includegraphics[width=3.29in]{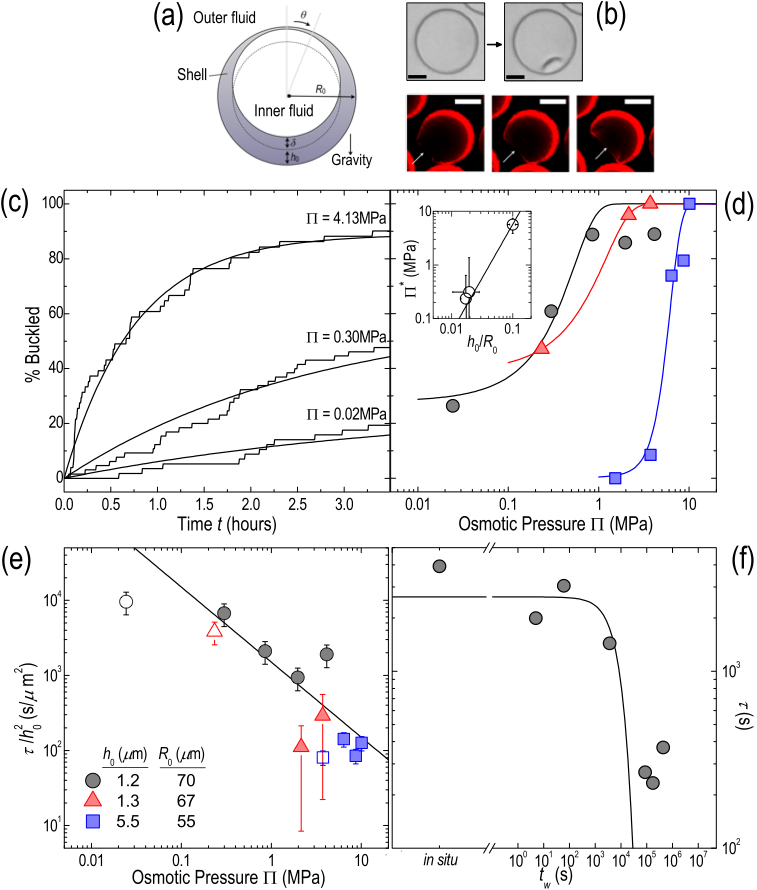}
\caption{(a) Schematic showing the capsule geometry investigated. (b) Upper: buckling of a capsule; scale bar is $20\mu$m. Lower: buckling begins at the thinnest part of the shell for capsules with thickness inhomogeneity $\delta/h_{0}\approx0.84$; scale bars are $50\mu$m. (c) Fraction of capsules buckled over time, for three different osmotic pressures $\Pi$. Capsules have mean shell thickness $h_{0}=1.2\mu$m, outer radius $R_{0}=70\mu$m, and $\delta/h_{0}=0.20$. Smooth lines show exponential fits. (d) Total fraction of capsules that ultimately buckle over time for varying $\Pi$, for capsules with $h_{0}, R_{0},$ and $\delta/h_{0}=1.2\mu$m, $70\mu$m, and $0.20$ (grey circles), $1.3\mu$m, $67\mu$m, and $0.23$ (red triangles), and $5.5\mu$m, $55\mu$m, and $0.19$ (blue squares). Smooth curves are fits to the data using the cumulative distribution function of the normal distribution. Inset shows mean osmotic pressure of each fit versus $h_{0}/R_{0}$; vertical and horizontal error bars show standard deviation of each fit and estimated variation in $h_{0}/R_{0}$, respectively. Straight line shows $(h_{0}/R_{0})^{2}$ scaling. (e) Time delay before the onset of buckling, $\tau$, normalized by $h_{0}^{2}$, for varying $\Pi$, for the same capsules as in (d). Filled points show $\Pi>\Pi^{*}$ while open points show $\Pi<\Pi^{*}$. Vertical error bars show uncertainty arising from estimated variation in $h_{0}$. Black line shows $\Pi^{-1}$ scaling. (f) Time delay $\tau$ decreases with the wait time before a shell is polymerized, $t_{w}$; capsules have $h_{0}=1.2\mu$m and $R_{0}=70\mu$m, and are buckled at $\Pi\approx0.86$ MPa $>\Pi^{*}$. Black line shows theoretical prediction coupling shell theory and Darcy's law, as described in the text, with $k\approx3.5\times10^{-24}$ m$^{2}$.}  
\end{center}
\end{figure}

The osmotic pressure difference across these inhomogeneous shells forces the capsules to buckle; we observe the abrupt formation of localized indentations in the shells, as shown in Fig. 1(b). For each osmotic pressure investigated, the fraction of the capsules that buckle increases over time, eventually plateauing, as shown in Fig. 1(c). We quantify this behavior by fitting this increase to an empirical exponential relationship, exemplified by the smooth lines in Fig. 1(c). The plateau value of this function yields a measure of the total fraction of the capsules that ultimately buckle over sufficiently long times, while the time constant of this function yields a measure of the time delay before the onset of buckling, $\tau$. For sufficiently large $\Pi$, the total fraction of the capsules that ultimately buckle increases dramatically with increasing $\Pi$, as shown by the grey circles in Fig. 1(d); this indicates that the capsules buckle above a threshold pressure, $\Pi^{*}$ \cite{jayson1}. We empirically fit these data using the cumulative distribution function of a normal distribution, shown by the black line in Fig. 1(d); the mean value and standard deviation of this fit yield a measure of $\Pi^{*}$ and the spread in $\Pi^{*}$, respectively. 

We study the geometry dependence of $\Pi^{*}$ by performing additional measurements on inhomogeneous capsules, polymerized {\it in situ}, with different shell thicknesses and radii; these are characterized by $\delta/h_{0}\approx0.2$, and $h_{0}/R_{0} = 0.019$ or $h_{0}/R_{0}=0.1$. Similar to the $h_{0}/R_{0}=0.017$ case, for sufficiently large $\Pi$, the total fraction of the capsules that ultimately buckle increases dramatically with increasing $\Pi$, as shown by the red triangles and blue squares in Fig. 1(d). Interestingly, we find that the threshold buckling pressure $\Pi^{*} \sim (h_{0}/R_{0})^{2}$ [Fig. 1(d), inset]; this observation is reminiscent of the prediction of shell theory for the buckling of a {\it uniform} shell \cite{koiter}, despite the fact that our capsules are inhomogeneous. To understand this behavior, we consider the local deformability of an inhomogeneous shell at various points on its surface. Because the 2D stretching and bending stiffnesses scale as $\sim h$ and $\sim h^{3}$ \cite{landau}, respectively, the thinnest part of the shell, where $h \approx h_{0}-\delta$, should be the easiest to deform. We directly visualize that buckling begins at this ``weak spot" using confocal microscopy of inhomogeneous capsules with fluorescent shells characterized by $\delta/h_{0}\approx0.84$ [Fig. 1(b), lower panel]. Consequently, we expect the onset of buckling to be governed by deformations in this part of the shell. To quantify this expectation, we apply shell theory to an inhomogeneous shell characterized by the same geometry as the experimental capsules \cite{supp}; this analysis yields $\Pi^{*}=\frac{2E}{\sqrt{3(1-\nu^{2})}} \left(\frac{h_{0}-\delta}{R_{0}}\right)^{2} \approx470(h_{0}/R_{0})^{2}$ MPa, assuming a Poisson ratio $\nu\approx 1/3$. The dependence of $\Pi^{*}$ on $h_{0}-\delta$ confirms our expectation that the threshold buckling pressure is set by the thinnest part of the inhomogeneous shell. Moreover, we find $\Pi^{*}(R_{0}/h_{0})^{2}\approx600\pm200$ MPa for the experimental capsules [solid line, Fig. 1(d) inset], in good agreement with our theoretical prediction. This indicates that the onset of capsule buckling is well described by shell theory. 

Within this framework, for $\Pi>\Pi^{*}$, a capsule remains spherical before it buckles; it initially responds to the applied pressure by contracting uniformly, reducing its volume from its initial value, $V_{0}$, by a threshold amount $\Delta V^{*}$, before buckling. We find that the time delay before the onset of buckling, $\tau$, strongly decreases with increasing osmotic pressure $\Pi>\Pi^{*}$, as shown by the filled points in Fig. 1(e). We hypothesize that this behavior reflects the dynamics of the fluid flow through the capsule shell; for the capsule to buckle, a volume $\Delta V^{*}$ of fluid must be ejected from its interior. The time delay can then be estimated as $\tau=\Delta V^{*}/Q$ \cite{stress}, where both $\Delta V^{*}$ and $Q$, the volumetric rate of fluid ejection from the capsule interior, are functions of $\delta/h_{0}$. We calculate $\Delta V^{*}$ for inhomogeneous shells using shell theory and validate the calculations with numerical simulations; the fluid ejection rate $Q$ follows from integrating Darcy's law over the surface of the capsule geometry shown in Fig. 1(a) \cite{supp}. Combining these results, we obtain 
\begin{equation} \label{eqn-tau}
\tau\approx\frac{V_{0}}{Q_{0}}\sqrt{\frac{3(1-\nu)}{1+\nu}}\frac{h_{0}}{R_{0}}\left(1-\frac{\delta}{h_{0}}\right)^{2}
\end{equation}
where $Q_{0}\equiv4\pi R_{0}^{2}\Pi k/\mu h_{0}$ and $k$ is the shell permeability. For the inhomogeneous capsules, characterized by $\delta/h_{0}\approx0.2$, we thus expect $\tau/h_{0}^{2}\approx 0.8\mu/k\Pi$; our experimental measurements of $\tau$ allow a direct test of this prediction. Above $\Pi^{*}$, the data collapse when $\tau$ is rescaled by $h_{0}^{2}$, as shown by the filled points in Fig. 1(e), consistent with our expectation; moreover, by fitting these data [black line in Fig. 1(d)], we obtain an estimate for the shell permeability, $k\approx7\times10^{-24}$ m$^{2}$. We use optical microscopy to directly measure the rate at which the capsule volume decreases immediately after the onset of buckling \cite{supp}; this gives an independent measure of the shell permeability. We find $k\approx2\times10^{-24}$ m$^{2}$ [Fig. S8], in good agreement with the fit shown in Fig. 1(e); this further confirms the validity of Eq.~\ref{eqn-tau}. 

To test the applicability of this picture to even more inhomogeneous capsules, we measure $\tau$ for capsules polymerized at different $t_{w}$; these have shells with $h_{0}/R_{0}=0.017$ and $\delta/h_{0}$ ranging from 0.2 up to 0.84. We impose a fixed osmotic pressure $\Pi\approx0.86$ MPa $>\Pi^{*}$. We observe that $\tau$ decreases only slightly with increasing $t_{w}<10^{3}$ s; however, as $t_{w}$ is increased above this value, $\tau$ drops precipitously over one order of magnitude, as shown by the points in Fig. 1(f). To quantitatively compare these data to Eq. (\ref{eqn-tau}), we estimate the dependence of $\delta/h_{0}$ on $t_{w}$ using lubrication theory; we validate this calculation using direct measurements of $\delta/h_{0}$ for capsules prepared at varying $t_{w}$ \cite{supp,howard}. Remarkably, we find good agreement between our data and Eq. (1), with $k\approx3.5\times10^{-24}$ m$^{2}$, as shown by the black line in Fig. 1(f); in particular, this simple picture captures the strong decease in $\tau$ at $t_{w}\sim10^{3}$s, with a shell permeability consistent with our independent measurements [Fig. S8]. While these results do not rule out other possible functional forms of $\tau$, they further suggest that the time delay before the onset of buckling can be understood by combining shell theory with Darcy's law for flow through the capsule shell, even for very inhomogeneous shells. 

\begin{figure}
\begin{center}
\includegraphics[width=3.5in]{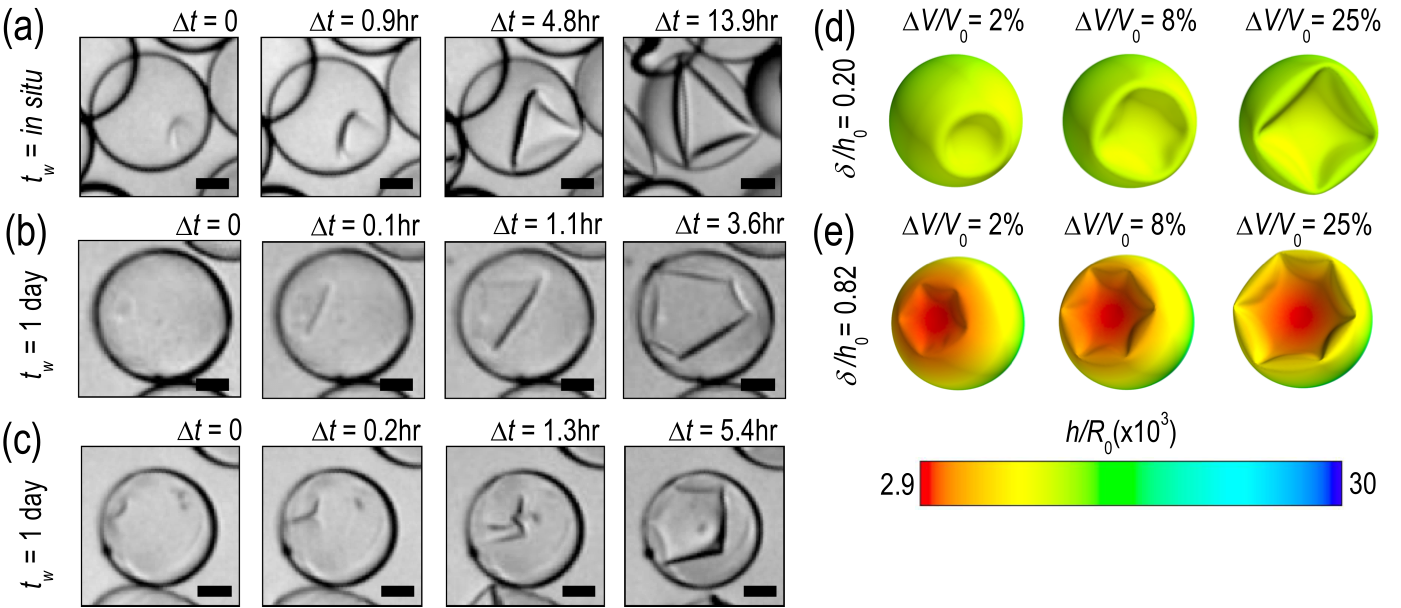}
\caption{Folding pathways for different shell inhomogeneities. (a-c) Optical microscope images exemplifying buckling at $\Pi\approx0.86$ MPa of (a) slightly inhomogeneous capsules polymerized {\it in situ} ($t_{w}\approx0$), with $\delta/h_{0}\approx0.2$, (b-c) very inhomogeneous capsules polymerized after a wait time $t_{w}=1$ day, with $\delta/h_{0}\approx0.84$. Very inhomogeneous capsules buckle through the formation of either (b) one single indentation or (c) two indentations. $\Delta t$ is time elapsed after buckling. Scale bars are $35\mu$m. (d-e) Examples of simulated shells with similar geometries as the capsules shown in (a-c), for varying fractional volume reduction $\Delta V/V_{0}$. Color scale indicates the spatially-varying shell thickness.} 
\end{center}
\end{figure}

The shell thickness inhomogeneity may continue to guide the development of deformations in a capsule after it buckles. To explore this possibility, we use optical microscopy to monitor the evolution of the capsule morphologies after the onset of buckling. Slightly inhomogeneous capsules typically buckle through the sudden formation of a single circular indentation. As this indentation grows over time, its perimeter eventually sharpens into straight ridges connected by 2-3 vertices \cite{vaziri, pauchard, vella2}; this folding pathway is exemplified by capsules polymerized {\it in situ}, characterized by $\delta/h_{0}\approx0.2$, as shown in Fig. 2(a). This sharpening reflects the unique physics of thin shells: because it is more difficult to compress the capsule shell than it is to bend it, localizing compressive deformations only along sharp lines and points on the capsule surface requires less energy than uniformly compressing the shell \cite{witten}. Interesting differences arise for very inhomogeneous capsules polymerized after $t_{w}=1$ day, characterized by $\delta/h_{0}\approx0.84$. The initial folding pathway is similar; however, the perimeters of the indentations formed in these capsules sharpen into straight ridges connected by 4-5 vertices, more than in the slightly inhomogeneous case, as shown in Fig. 2(b). Moreover, surprisingly, roughly $30\%$ of the very inhomogeneous capsules begin to buckle through the formation of one, then two, adjacent indentations, as exemplified in Fig. 2(c). The perimeters of these indentations grow over time, eventually meeting, coalescing, and sharpening into straight ridges connected by 4-5 vertices [Fig. 2(c)]. These observations directly demonstrate that the deformations of a capsule after it buckles are sensitive to the shell inhomogeneity.

To gain insight into this behavior, we perform numerical simulations of two different shells, a slightly inhomogeneous shell with $\delta/h_{0} = 0.20$, and a very inhomogeneous shell with $\delta/h_{0} = 0.82$, similar to the experimental capsules. As the shell volume is reduced below $V_{0}-\Delta V^{*}$, both shells buckle through the formation of a single indentation centered at the thinnest part of the shell, as shown in the leftmost panels of Fig. 2(d-e). As $\Delta V$ increases, this indentation grows and its edges sharpen. We find that the indentations formed in the very inhomogeneous shells begin to sharpen at smaller $\Delta V/V_{0}$, and ultimately develop more vertices than those formed in more homogeneous shells [Fig. 2(d-e)] \cite{supp}. These results qualitatively agree with our experimental observations [Fig. 2(a-c)], further confirming that after the onset of buckling, the folding pathway of a shell depends on the inhomogeneity. However, in contrast to the experimental capsules [Fig. 2(c)], we do not systematically observe the formation of adjacent indentations in the simulations on very inhomogeneous shells \cite{double}. This presents a puzzle requiring further inquiry.

\begin{figure}
\begin{center}
\includegraphics[width=2.35in]{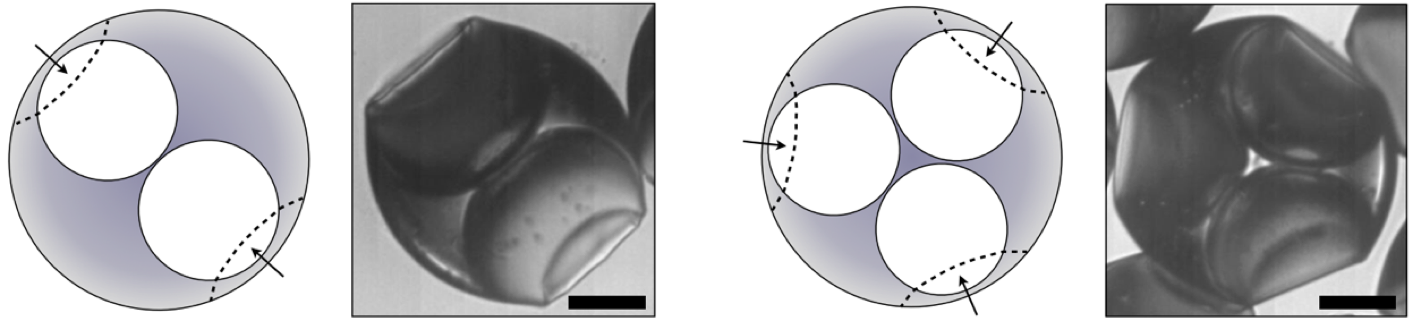}
\caption{Colloidal capsules with two or three spherical interior compartments, schematized in left panels, buckle at ``weak spots" (arrows). This forms shapes with two or three equally-spaced circular indentations after buckling (right panels). Scale bars are $100\mu$m.} 
\end{center}
\end{figure}

Our capsules may be used to guide colloidal self-assembly; for example, a colloidal particle can spontaneously bind to the indentation formed during buckling through a lock-and-key mechanism \cite{sacanna}. This mechanism is typically applied to a homogeneous colloidal particle, which buckles through the formation of a single indentation at a random position on its surface. We apply our findings to create multiply-indented capsules having two-fold or three-fold symmetry. To do this, we form double emulsions with two or three inner droplets of radii larger than half the radius of the outer droplet. Consequently, the inner droplets pack closely to form dimers or trimers \cite{shin3}, as shown schematically in Fig. 3. The double emulsions are then polymerized, forming solid capsules with two or three spherical compartments in their interiors, and two or three equally-spaced ``weak spots" in the capsule shell [arrows in Fig. 3]. When exposed to a sufficiently large osmotic pressure, these capsules buckle through the formation of multiple, equally-spaced indentations at the weak spots, as shown in Fig. 3. This approach is thus a versatile way to create capsules of desired symmetries, and extends the range of structures that can be used for lock-and-key colloidal assembly. \\

\noindent This work was supported by Amore-Pacific, the NSF (DMR-1006546) and the Harvard MRSEC (DMR-0820484). Work by DRN was also supported by NSF grant DMR-1005289. SSD acknowledges funding from ConocoPhillips. It is a pleasure to acknowledge H. A. Stone for communicating the lubrication theory solution used here; D. Vella for stimulating discussions and for helpful feedback on the manuscript; G. M. Whitesides for comments that motivated some of this work; and the anonymous referees for valuable feedback on the manuscript. The computer programs used to conduct numerical simulations are based on code generously provided by E. Katifori.


\begin{thebibliography}{10}
\bibitem{spread}
Each batch is characterized by a range of waiting times, with an average value $t_{w}$ \cite{supp}.
\bibitem{sem}
We measure the capsule geometric properties using either SEM or confocal microscopy of fluorescent capsules.
\bibitem{stress}
The fluid ejection is resisted by the compressive stress in the shell; however, this stress can be omitted from Darcy's law for the range of $\Pi$ studied here (\cite{supp} and D. Vella, private communication, March 2012).
\bibitem{double}
We observe this phenomenon for the inhomogeneous capsules buckled over a range of $\Pi\approx0.5-5$ MPa, and do not observe it for the more homogeneous shells buckled at large pressures $\Pi\approx10$ MPa; this suggests that it is not simply a function of the rate at which the capsule volume is reduced \cite{gompper}.
\end{thebibliography}
	\end{document}

\title{Supporting Information: \\Delayed Buckling and Guided Folding of Inhomogeneous Capsules}

    \author{Sujit S. Datta}
    \affiliation{These authors contributed equally to this work.}
    \affiliation{Department of Physics, Harvard University, Cambridge MA 02138, USA}           

    \author{Shin-Hyun Kim}
    \affiliation{These authors contributed equally to this work.}
        \affiliation{Harvard School of Engineering and Applied Sciences, Cambridge MA 02138, USA}
  \affiliation{Department of Chemical and Biomolecular Engineering, KAIST, Daejeon, South Korea}

    \author{Jayson Paulose}
        \affiliation{These authors contributed equally to this work.}
    \affiliation{Harvard School of Engineering and Applied Sciences, Cambridge MA 02138, USA}

    \author{Alireza Abbaspourrad}
    \affiliation{Harvard School of Engineering and Applied Sciences, Cambridge MA 02138, USA}
    \author{David R. Nelson}
    \affiliation{Department of Physics, Harvard University, Cambridge MA 02138, USA}
    \author{David A. Weitz}
    \affiliation{Department of Physics, Harvard University, Cambridge MA 02138, USA}
    \affiliation{Harvard School of Engineering and Applied Sciences, Cambridge MA 02138, USA}

\maketitle

\section{Microfluidic fabrication of capsules}
To prepare monodisperse capsules, we use W/O/W double-emulsion droplets made with two different types of glass capillary microfluidic devices. To make capsules with $h_{0}/R_{0}=0.017$, we use a device composed of a hydrophobic tapered injection capillary (World Precision Instruments, Inc., 1B100-6), pre-treated with n-octadecyltrimethoxyl silane (Aldrich), inserted in a second square capillary (AIT glass) \cite{shin}; the inner diameter of the square capillary and the outer diameter of the injection capillary are both 1 mm. Furthermore, a small tapered capillary is inserted into the injection capillary to simultaneously inject two immiscible fluids, as shown in Fig. S1(a). To confine the flow near the injection tip and thereby increase the flow velocity, a hydrophilic circular capillary, pre-treated with 2-[methoxy(polyethyleneoxy)propyl] trimethoxyl silane (Gelest, Inc.), is inserted into the square capillary at its other end.

To make W/O/W double-emulsion droplets, we simultaneously introduce two immiscible phases, 10 wt\% aqueous solution of poly (vinylalcohol) (PVA, Mw 13 000-23 000) and ethoxylated trimethylolpropane triacrylate (ETPTA) containing 0.2 wt\% photoinitiator (2-hydroxy-2-methylpropiophenone, Aldrich), through the injection capillary at flow rates of 300 $\mu$L/h and 150 $\mu$L/h, respectively. The aqueous solution is saturated with ETPTA before injection to prevent diffusion of ETPTA molecules through water. We use the same aqueous solution of PVA as the continuous phase; this is injected through the square capillary at the flow rate of 3000 $\mu$L/h. Because of the hydrophobic nature of the injection capillary, the ETPTA flows along the inner surface of the injection capillary; by contrast, the aqueous solution flows through the center of the injection capillary as plug-like droplets that do not contact the wall. These plug-like droplets are emulsified at the tip of the injection capillary, resulting in monodisperse double-emulsion droplets with an ultra-thin middle layer as shown in Fig. S1(b). The ETPTA formed between plug-like droplets produces large blobs at the tip of the injection capillary; we separate these from the double-emulsion droplets by exploiting their density difference. We prepare thin-shelled capsules by photo-polymerization of the droplet middle phase, ETPTA. 

To make capsules with $h_{0}/R_{0}=0.019$ and 0.1, we use a device composed of hydrophobic injection and collection capillaries as shown in Fig. S2(a) \cite{andy}. The innermost aqueous phase is injected through the injection capillary and the middle ETPTA phase is injected through the interstices of the injection and the square capillaries. The continuous aqueous phase is injected through the interstices of the collection and the square capillaries as a counter flow to the innermost and the middle phases. These three streams flow coaxially through the orifice of the collection capillary, making double-emulsion drops in a dripping mode. For $h_{0}/R_{0}=0.019$, flow rates of the innermost, the middle, and the continuous phases are kept at 400, 80, 2000 $\mu$L/h, respectively, as shown in Fig. S2(b). For $h_{0}/R_{0}=0.1$, flow rates of the innermost, the middle, and the continuous phases are kept at 200, 100, 2000 $\mu$L/h, respectively, as shown in Fig. S2(c).

Capsules photopolymerized {\it in situ} are prepared in the following manner: As soon as the drops flow out from the nozzle of a microfluidic device to water bath containing 8 wt\% aqueous solution of PVA which is saturated with ETPTA, they are polymerized by continuous UV irradiation (Omnicure S1000). 

Capsules with $t_{w} = 5$ s or 1 min are prepared in the following manner: As they are continuously produced from the nozzle of the microfluidic device, the droplets are collected into a chamber containing 8 wt\% aqueous solution of PVA saturated with ETPTA and are repetitively irradiated with UV irradiation (Omnicure S1000) of duration 1 second, fully polymerizing the ETPTA into a thin solid shell, every $2\times t_{w}$. This forms a population of capsules that have remained quiescent in the collection chamber for an average waiting time of $t_{w}$ before polymerization.

Capsules with $t_{w} = 1$ hr, 1 day, 2 days or 5 days are prepared in the following manner: As they are continuously produced from the nozzle of the microfluidic device, the droplets are collected for a total time of 10 min into a chamber containing 10 wt\% aqueous solution of PVA saturated with ETPTA. They are then photopolymerized after a waiting time of $t_{w} - 5$min.

Optical microscope images of monodisperse microcapsules with $t_{w} = 1$ min are shown in Fig. S3.

To make microcapsules with two or three spherical compartments, we employ two-step emulsification in a capillary microfluidic device \cite{shin1}. Inner water drops are produced at the tip of small tapered capillary in a dripping mode, which in turn are encapsulated into oil drops at the tip of injection capillary. Through precise control of flow rates of three streams, we can manipulate generation frequency of inner water drops and outer oil drops. Therefore, the number of inner drops confined in outer drop is controllable. To make the thin shell, the flow rate of middle oil stream is maintained to be small enough to encapsulate two or three inner drops with nonspherical envelope. The nonspherical double-emulsion drops are polymerized by {\it in-situ} UV irradiation, resulting in microcapsules with multiple spherical compartments.

To buckle the capsules, we gently mix the capsule suspension with the NaCl solution by repeatedly aspiring and ejecting the fluid with a micro-pipette; this ensures the capsules are more directly exposed to the external osmotic pressure. 

We use either scanning electron microscopy (SEM) or confocal microscopy to directly measure the capsule geometrical characteristics.

\begin{figure}[h]
\begin{center}
\includegraphics[width=3.0in]{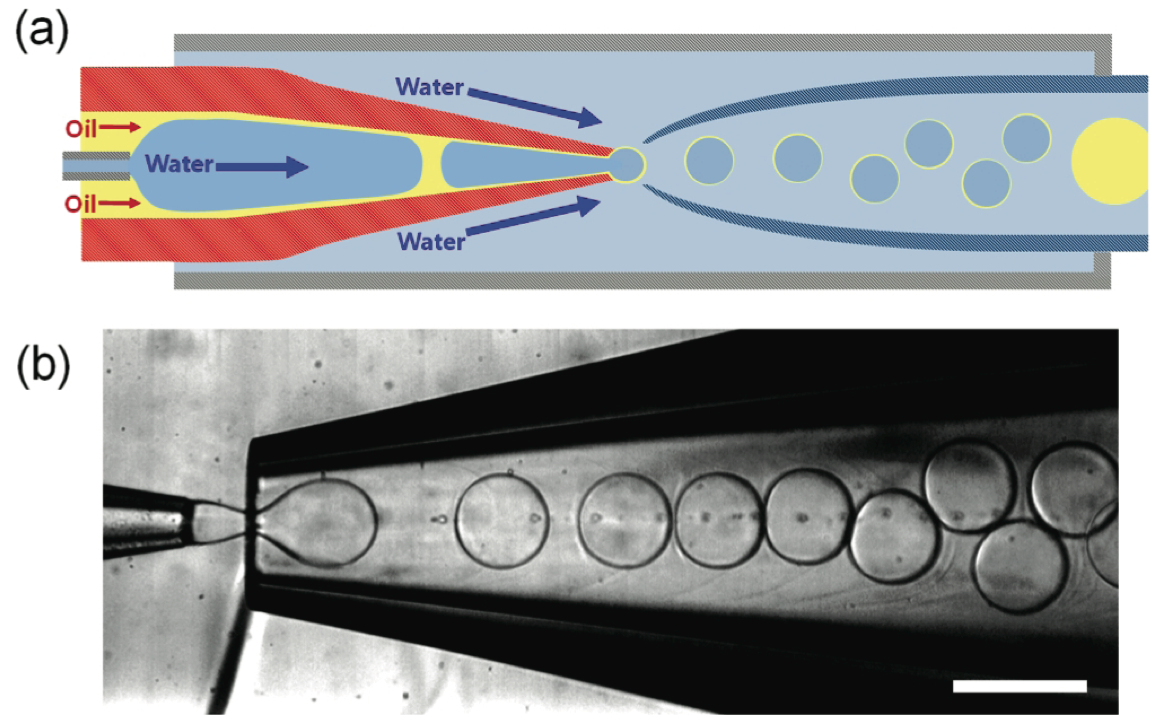}
\caption{Microfluidic preparation of double-emulsion drops with $h_{0}/R_{0}=0.017$. Scale bar is 200$\mu$m. } 
\end{center}
\end{figure}

\begin{figure}[h]
\begin{center}
\includegraphics[width=3.0in]{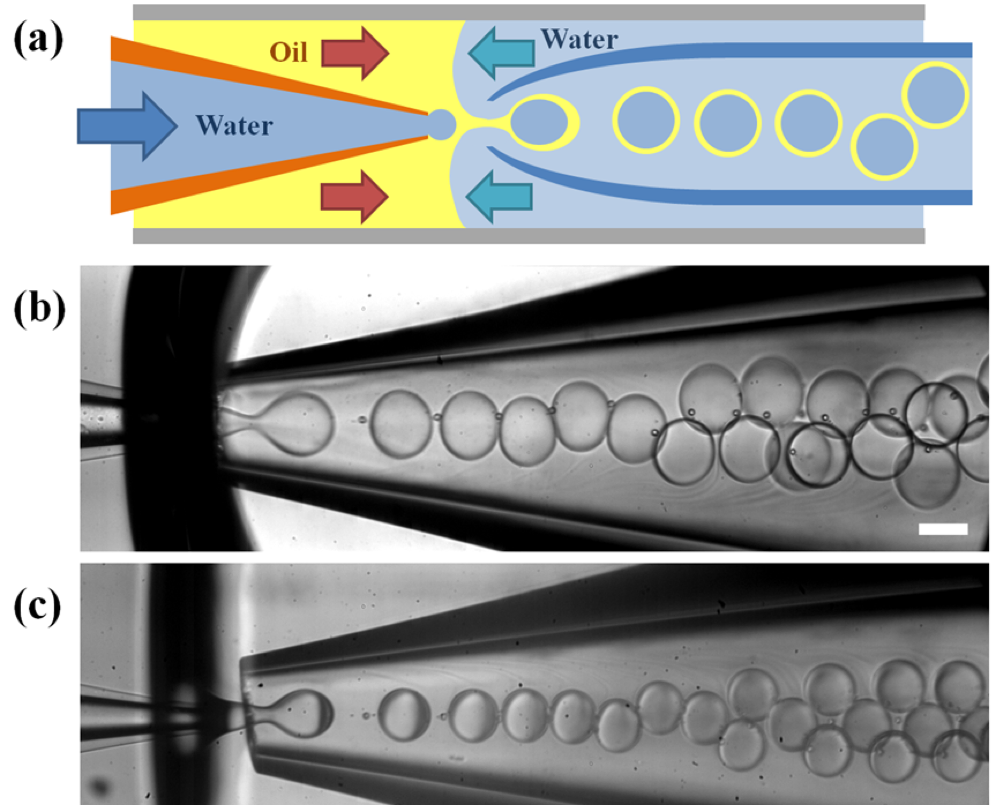}
\caption{Microfluidic preparation of double-emulsion drops with (b) $h_{0}/R_{0}=0.019$ and (c) $h_{0}/R_{0}=0.1$. Scale bars are 100$\mu$m. } 
\end{center}
\end{figure}

\begin{figure}[h]
\begin{center}
\includegraphics[width=3.0in]{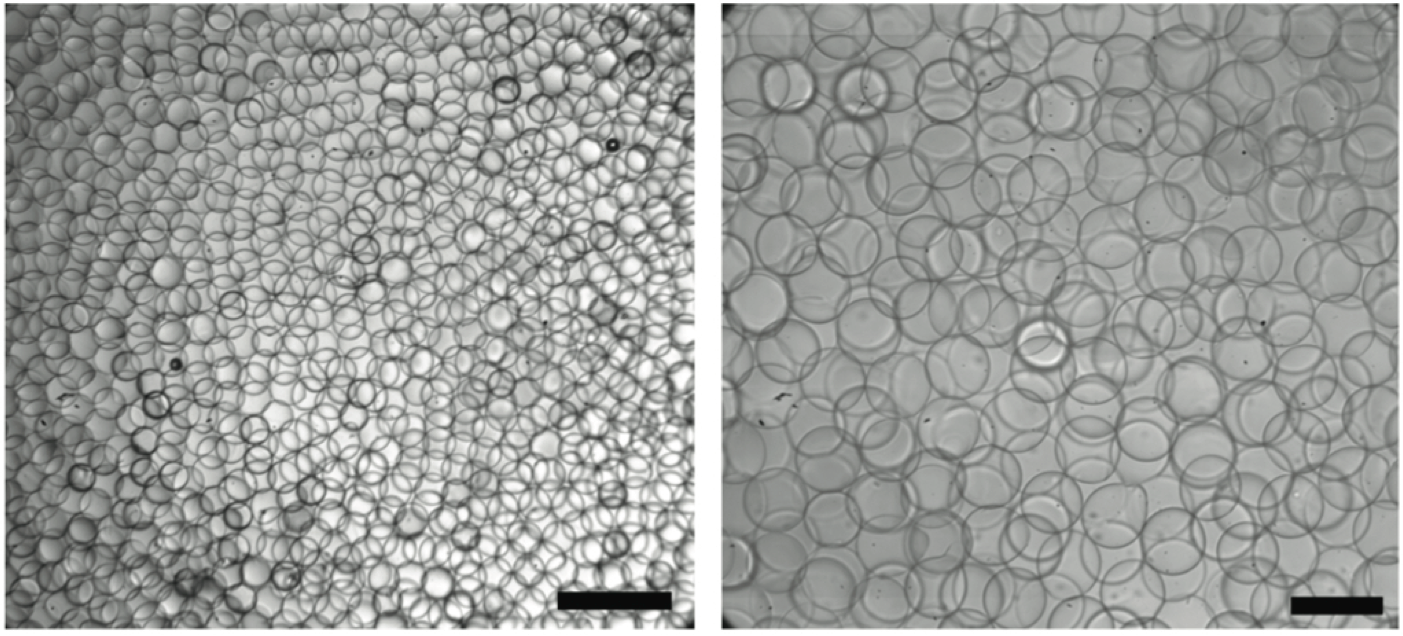}
\caption{Monodisperse microcapsules. Scale bars in (a) and (b) are 500 $\mu$m and 200 $\mu$m, respectively.} 
\end{center}
\end{figure}

\begin{figure}[h]
\begin{center}
\includegraphics[width=3.0in]{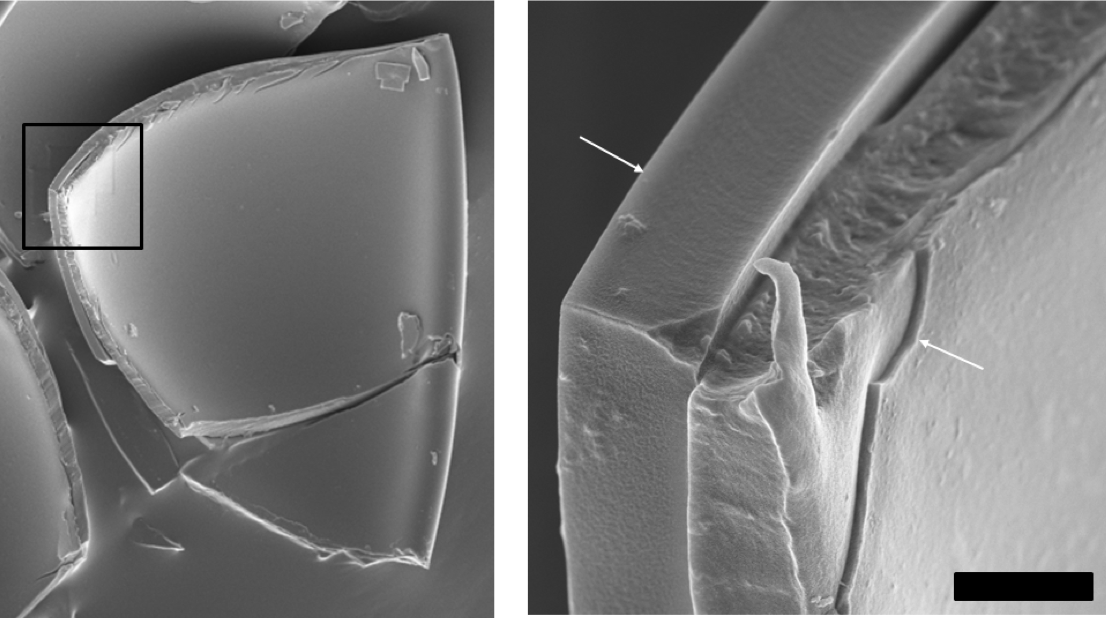}
\caption{SEM showing buckled shell with inhomogeneity. Shells buckle at the thinnest part; as a result, the thinnest part contacts the thickest part of the shell. Right panel: the left arrow shows the thickest part of the shell, and the right arrow shows the thinnest part of the shell; the solid in between the two parts of the shell is precipitated PVA. Scale bar is $2\mu$m.} 
\end{center}
\end{figure}

\newpage
\section{Buckling theory of inhomogeneous shells}
\begin{figure}[h]
    \includegraphics[width=3in]{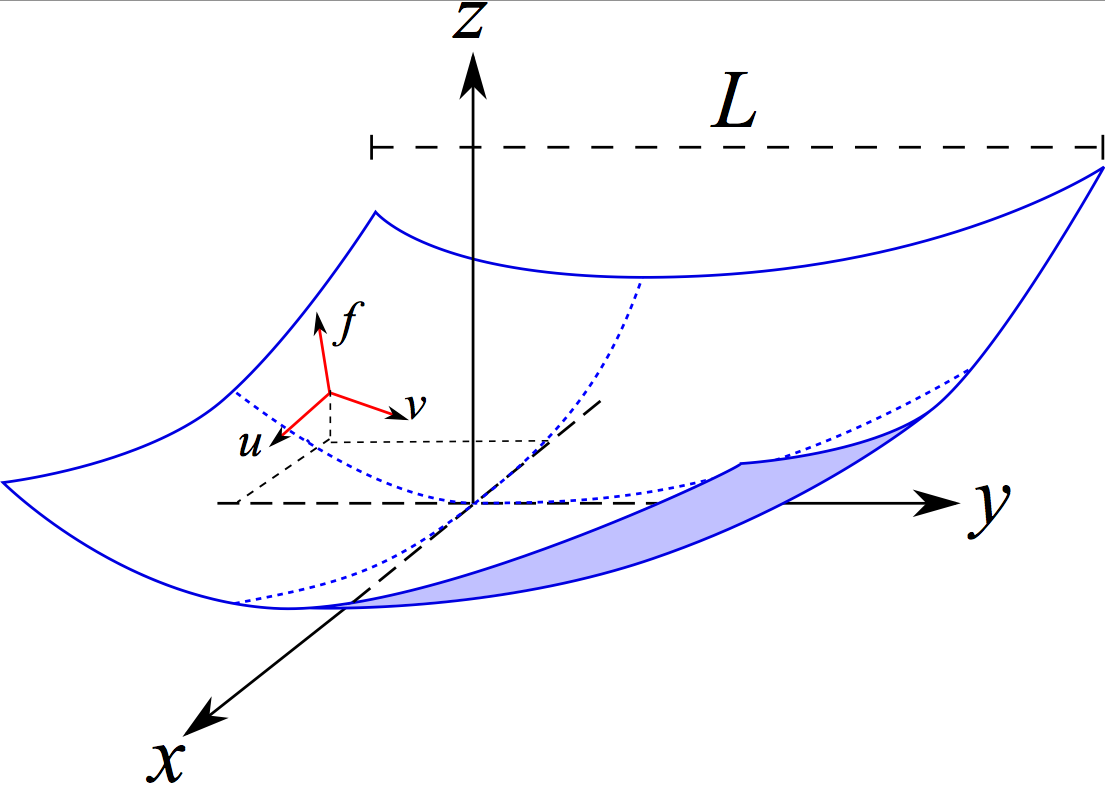}
    \caption{Coordinates and displacements in shallow shell theory.} \label{fig-shallow}
\end{figure}

Before we derive the threshold buckling pressure $\Pi^{*}$ and associated volume change $\Delta V^{*}$ for inhomogeneous shells, it is instructive to study the derivation for homogeneous shells of thickness $h$, elastic modulus $E$, Poisson's ratio $\nu$ and radius $R$. We use \emph{shallow-shell theory} and follow the presentation by Hutchinson (ref.~\onlinecite{hutchinson}). The shallow-shell description involves isolating a shallow section of the shell of size $ \sim L$ and defining a Cartesian coordinate system $(x,y)$ tangential to it (Fig.~\ref{fig-shallow}). ``Shallowness'' refers to the condition that the section is small compared to the radius, $L/R \ll 1$, so that slopes of the surface are small; the validity of this approach for the buckling solution will be checked later. The shape of the middle surface is described by in-plane displacement fields $u,v$ and a transverse displacement $w$ of the middle surface of the shell. Ignoring terms of order $(x/R)^{2},(y/R)^{2}\ll 1$, the displacement-strain relations for the section are
\begin{multline} \label{eqn_strains}
%~ \begin{split}
\gamma_{xx} = u_{,x}-\frac{w}{R}+\frac{1}{2}w_{,x}^2, \quad\\
\gamma_{xy} = \frac{1}{2}\left(u_{,y}+v_{,x}+w_{,x}w_{,y}\right), \quad\\
\gamma_{yy} = v_{,y} - \frac{w}{R}+\frac{1}{2}w_{,y}^2.
%~ \end{split}
\end{multline}

The elastic energy of the shell involves stretching, bending and a pressure term:
\begin{multline} \label{eqn0etot}
E_\mathrm{tot}[\mathbf{r}]=\\
\iint\left[\frac{1}{2}\gamma_{ij}\sigma_{ij}+\frac{Eh^3}{24(1-\nu^2)}(w_{,xx}+w_{,yy})^2 -\Pi w\right]\,dx\,dy
\end{multline}
where the stresses $\sigma_{ij}$ are related to the strains via
\begin{multline} \label{eqn_stresses}
%~ \begin{split}
\gamma_{xx} = \frac{1}{Eh}(\sigma_{xx} -\nu \sigma_{yy}), \quad\\
\gamma_{yy} = \frac{1}{Eh}(\sigma_{yy} -\nu \sigma_{xx}), \quad\\
\gamma_{xy} = \frac{1+\nu}{Eh}\sigma_{xy}. 
%~ \end{split}
\end{multline}

The initial response of the shell to an external pressure $\Pi$ is to contract uniformly, building up uniform compressive shell stresses 
\begin{equation} \label{eqn0sigma0}
    \sigma_{xx}^{0}=\sigma_{yy}^{0}= -\Pi R/2
\end{equation}
with corresponding transverse displacement field
\begin{equation} \label{eqn0w0}
    w_{0} = \frac{\Pi R^{2}(1-\nu)}{2Eh}.
\end{equation}
that is also uniform over the whole section. The elastic energy associated with this compressed state, which does not involve bending of the shell surface, is roughly $\sigma^{2}/(Eh)\sim \Pi^{2}R^{2}/(Eh)$ per unit area. When $\Pi$ gets large, it becomes energetically favorable for the shell to introduce some bending deformations to trade the elastic energy of pure compression for bending energy. The buckling transition is identified by the value of $\Pi$ for which a nonuniform deformation mode of the shell first arises which is energetically favorable compared to the uniform compression. A linear buckling analysis of the Euler-Lagrange equations associated with the energy functional (Eq.~\ref{eqn0etot}) shows that a nontrivial solution for the transverse displacement field $w$ first arises at a threshold buckling pressure~\cite{hutchinson}
\begin{equation} \label{eqn0pcrituniform}
    \Pi^{*} = \frac{2E}{\sqrt{3(1-\nu^{2})}} \left(\frac{h}{R}\right)^{2}.
\end{equation}
The form of the associated displacement field is sinusoidally, $w = A e^{i\mathbf{q}\cdot\mathbf{x}},$ with period set by a two-dimensional wavevector $\mathbf{q}$ that satisfies $|\mathbf{q}| = q^{*} \equiv [12(1-\nu^2)]^{1/4}/\sqrt{hR}$. (The corresponding solutions for the in-plane displacement fields $u$ and $v$ have similar forms.) Since a multitude of two-dimensional wavevectors exists with magnitude equal to $q^{*}$, there are many degenerate buckling modes at the transition. Crucially, the wavelength associated with the buckling modes is extremely small for thin shells:
\begin{equation}
    \lambda^{*} = 2\pi/q^{*} = \frac{2\pi}{[12(1-\nu^2)]^{1/4}}\sqrt{hR} \ll R.
\end{equation}
This observation justifies the use of shallow-shell theory to calculate the buckling pressure---a shallow section of the shell with $L/R \ll 1$ can still accommodate many wavelengths of the buckling modes and provides a good description of the elastic energy of these modes. 

We now consider an inhomogeneous shell with a continually varying shell thickness $h(\theta)=h_{0}-\delta\cos\theta$ and radius $R$ of the middle surface. By analogy with the buckling of homogeneous shells, we expect the buckling transition to be signalled by the existence of a nontrivial solution to the energy-minimizing displacement fields of the middle surface. Furthermore, we expect such a mode to first appear in the vicinity of the thinnest point of the sphere, $\theta = 0$. We consider a shallow section of the sphere, with Cartesian coordinates whose origin coincides with the thinnest point. As before, shallowness implies that the section is small enough that slopes are small compared to the base of the section. The elastic energy and the stress-strain relations are modified because the thickness of the shell is now position-dependent: $h(x,y) = h_{0}-\delta \left( 1-Z(x,y)/R\right),$ where $Z(x,y) = \sqrt{R^{2}-x^{2} -y^{2}}$ is the shape of the undeformed middle surface of the shell. Expanding in powers of $x/R$ and $y/R$, we have
\begin{equation}
    h(x,y) = h_{0}-\delta\left[1+ \frac{1}{2}\left(\frac{x^{2}+y^{2}}{R^{2}}\right) +...\right]
\end{equation}
but the shallowness of the section implies that $(x/R)^{2},(y/R)^{2}\ll 1$ and the section is well-described by setting $h \approx h_{0}-\delta$, a constant. Thus, the only modification to the shallow-shell theory calculation of the critical buckling pressure of the uniform shell is to set the shell thickness to $h = h_{0}-\delta$, from which we readily obtain the buckling pressure
\begin{equation} \label{eqn0pcritdelta}
\Pi^{*} = \frac{2E}{\sqrt{3(1-\nu^2)}}\left(\frac{h_{0}-\delta}{R}\right)^2.
\end{equation}
The wavelength of the buckling modes that arise at this pressure is $\lambda^{*} = 2\pi\sqrt{(h_{0}-\delta)R}/[12(1-\nu^2)]^{1/4} \ll R$, justifying the assumptions made. Essentially, the spatial thickness variation of the capsules under study is so gentle that there is hardly any variation in the mechanical properties of the capsule at the scale of the small wavelength associated with the buckling modes. Thus, the buckling pressure is set by the thickness of the capsule in the vicinity of the thinnest spot, where it is approximately $h_{0}-\delta$. (We could consider different sections of the capsule, with higher thicknesses, but the associated buckling mode would arise at a pressure higher than of Eq.~\ref{eqn0pcritdelta}.) Considering the magnitude of the terms excluded from the Euler-Lagrange equations by our approximation of constant thickness shows that the corrections to Eq.~\ref{eqn0pcritdelta} from taking into account the spatial variation in thickness over the shallow shell would be smaller by a factor $O(\delta/R)$. 

\begin{figure}[h]
\begin{center}
    \includegraphics[width=3.15in]{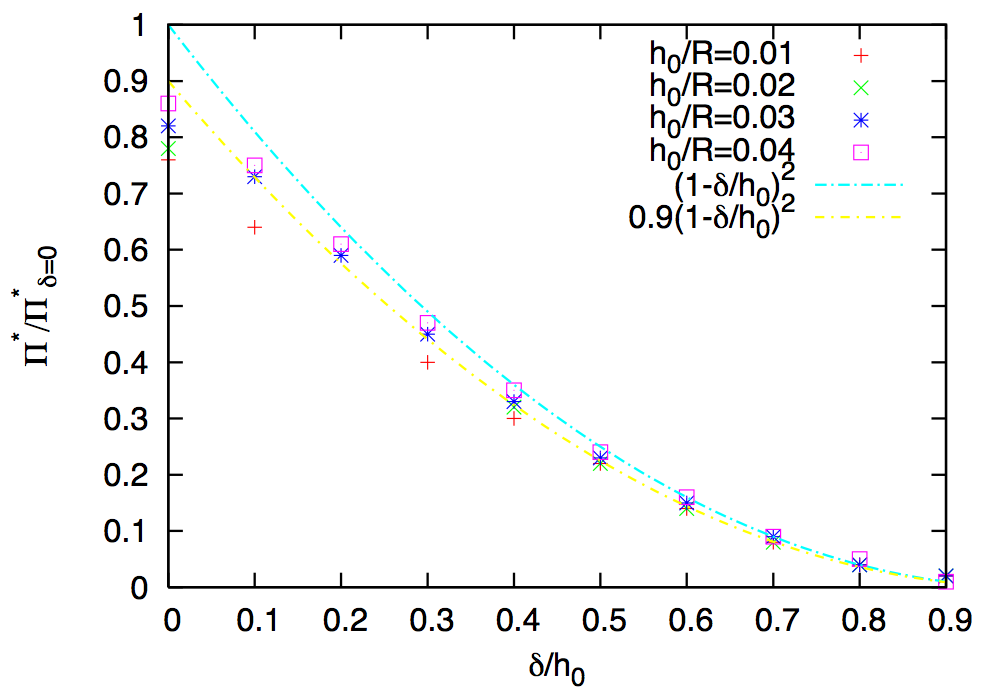} \\
    {\bf (a)}
    \includegraphics[width=3.15in]{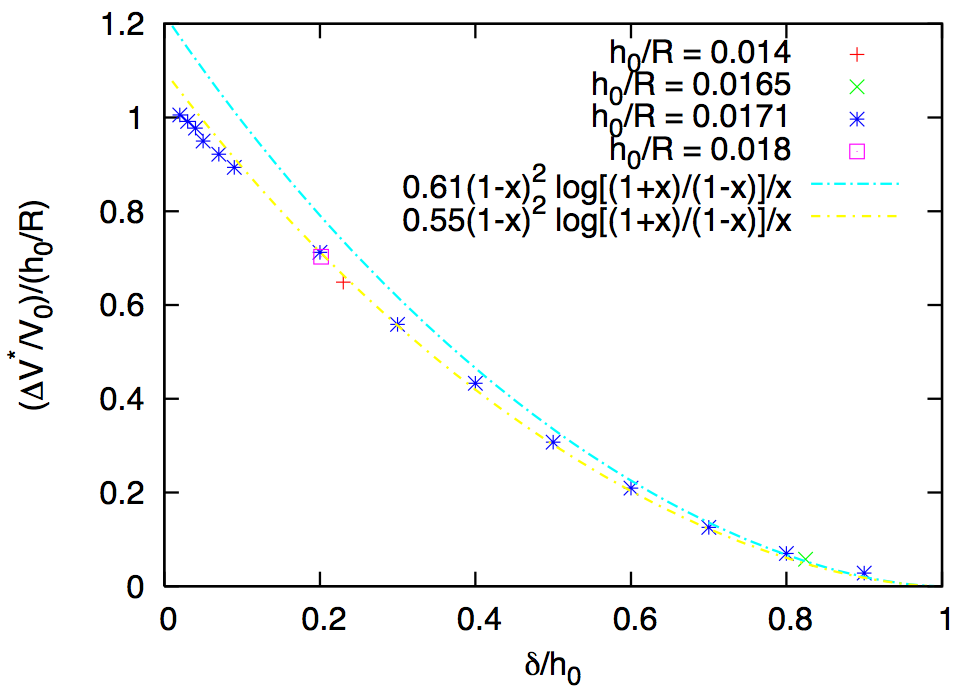} \\
    {\bf (b)}
\caption{{\bf (a)} Effect of inhomogeneity on the buckling pressure. The symbols show the buckling pressure from simulations for shells with various average thickness $h_{0}$ and inhomogeneity $\delta$, normalized by the corresponding buckling pressure for uniform shells with the same average thickness, $\Pi^{*}_{\delta=0} = 2E/\sqrt{3(1-\nu^{2})}\times(h_{0}/R)^{2}$ (Eq.~\ref{eqn0pcrituniform}). The blue line shows the theoretical prediction, Eq.~\ref{eqn0pcritdelta}, and the yellow line shows 90\% of the theoretical value. {\bf (b)} Threshold volume change at buckling, from simulations. The blue line shows the theoretical prediction (Eq.~\ref{eqn0vcdelta}), with $\nu = 1/3$, while the yellow line shows 90\% of the theoretical value. In both {\bf (a)} and {\bf (b)}, the theory agrees with the simulations to within about 90\%, and the theory effectively captures the dependence of the mechanical strength on the inhomogeneity. The systematic discrepancy of 10\% or so (higher for the extremely thin shells with $h_{0}/R = 0.01$) is likely a result of the sensitivity of the buckling transition to the small amount of disorder in the underlying mesh. It is known that the buckling transition of spherical shells is highly sensitive to imperfections in the shell~\cite{hutchinson} which significantly reduce the buckling pressure and through it the associated volume change.} \label{fig-PiStar}
\end{center}
\end{figure}

To test our theoretical arguments, we perform numerical simulations on shells with $0.01 \leq h_{0}/R \leq 0.04$ and $0 \leq \delta/h_{0} \leq 0.9$ (see section \ref{sec-simulations} below for details). The results are reported in Fig.~\ref{fig-PiStar}(a), showing good agreement between the theoretical prediction and the simulation results. The slight discrepancy between theory and simulation likely reflects the sensitivity of the buckling transition to non-uniformities in the simulation mesh. This sensitivity is  due to nonlinear couplings between the degenerate buckling modes that arise as the classically predicted buckling pressure is approached, which drive the transition to happen below the buckling pressure predicted by the linear stability analysis \cite{hutchinson}. We expect this sensitivity to be more pronounced for the uniform case, where the unstable modes extend over the entire shell, in contrast to the nonuniform case, where the unstable modes are localized in the shallow region near the thinnest part of the shell.

We now turn to the threshold volume change at buckling, $\Delta V^{*}$. The volume change prior to buckling is determined by the initial prebuckling response of the shell to the external pressure. For a spherical shell, the only axisymmetric stress distribution in response to a uniform pressure $p$ that is well-behaved at the poles is the uniform stress distribution, Eq.~(\ref{eqn0sigma0})~\cite{novozhilov}. In contrast to the uniform shell, however, the resulting transverse displacement varies with the shell thickness:
\begin{equation} \label{eqn0w0theta}
    w_{0}(\theta) =\frac{\Pi R^{2}(1-\nu)}{2Eh(\theta)} = \frac{\Pi R^{2}(1-\nu)}{2E(h_{0}-\delta cos\theta)}.
\end{equation}
The volume change in response to pressures up to the buckling pressure is thus (to lowest order in the inward displacement)
\begin{multline}
    %~ \begin{split}
        \Delta V = 2\pi\int_0^{\pi}R^{2}\sin\theta\,w_{0}(\theta )\,d\theta \\
        = \pi(1-\nu)\frac{\Pi R^{4}}{Eh_{0}}\left[\frac{h_{0}}{\delta}\ln\left(\frac{1+\delta/h_{0}}{1-\delta/h_{0}}\right)\right].
    %~ \end{split}
\end{multline}
The threshold volume change immediately before buckling is obtained by using $\Pi^{*}$ (Eqn.~\ref{eqn0pcritdelta}) in the above expression to get
\begin{multline} \label{eqn0vcdelta}
    \frac{\Delta V^{*}}{V_{0}}=\\
    \frac{1}{2}\frac{h_{0}}{R}\sqrt{\frac{3(1-\nu)}{1+\nu}}\left[\left(1-\frac{\delta}{h_{0}}\right)^{2}\frac{h_{0}}{\delta}\ln\left(\frac{1+\delta/h_{0}}{1-\delta/h_{0}}\right)\right]
\end{multline}
where $V_{0} = 4\pi R^{3}/3$ is the initial volume of the shell. We recover the result for uniform shells, $\Delta V^{*}/V_{0} = \sqrt{3(1-\nu)/(1+\nu)}\times h_{0}/R$ (Ref.~\onlinecite{knoche}) in the limit $\delta/h_{0}\to 0$. Fig.~\ref{fig-PiStar}(b) compares the analytical expression to numerical simulations on shells with various thicknesses and inhomogeneities, showing good agreement. Moreover, we find that the total fraction of capsules that ultimately buckle in our experiments increases with increasing $t_{w}$, and hence increasing inhomogeneity $\delta/h_{0}$, as shown in Fig. S7. This supports the theoretical finding that inhomogeneity reduces the mechanical strength of the shell.

We note that the linearized buckling analysis sketched out here can describe the shell up to the point of the buckling transition, but cannot describe the postbuckling shape of the shell. Once the buckling transition is reached, the nontrivial high-wavelength buckling mode, when it arises, coalesces into a single inversion (as shown, for example, in \cite{gompper}) unless an additional mechanism is introduced to arrest this coalescence Ð for example, by placing a slightly smaller, rigid, concentric sphere within the shell \cite{carlson}. In Carlson {\it et al}.'s work, the inner sphere prevented the unstable buckling mode from coalescing into a single inversion, giving rise instead to a regular dimple pattern. In the absence of such a stabilizing mechanism, the high-wavelength mode itself is not observed in buckling experiments including the ones on the inhomogeneous capsules prepared here.

\begin{figure}[h]
\begin{center}
\includegraphics[width=3.35in]{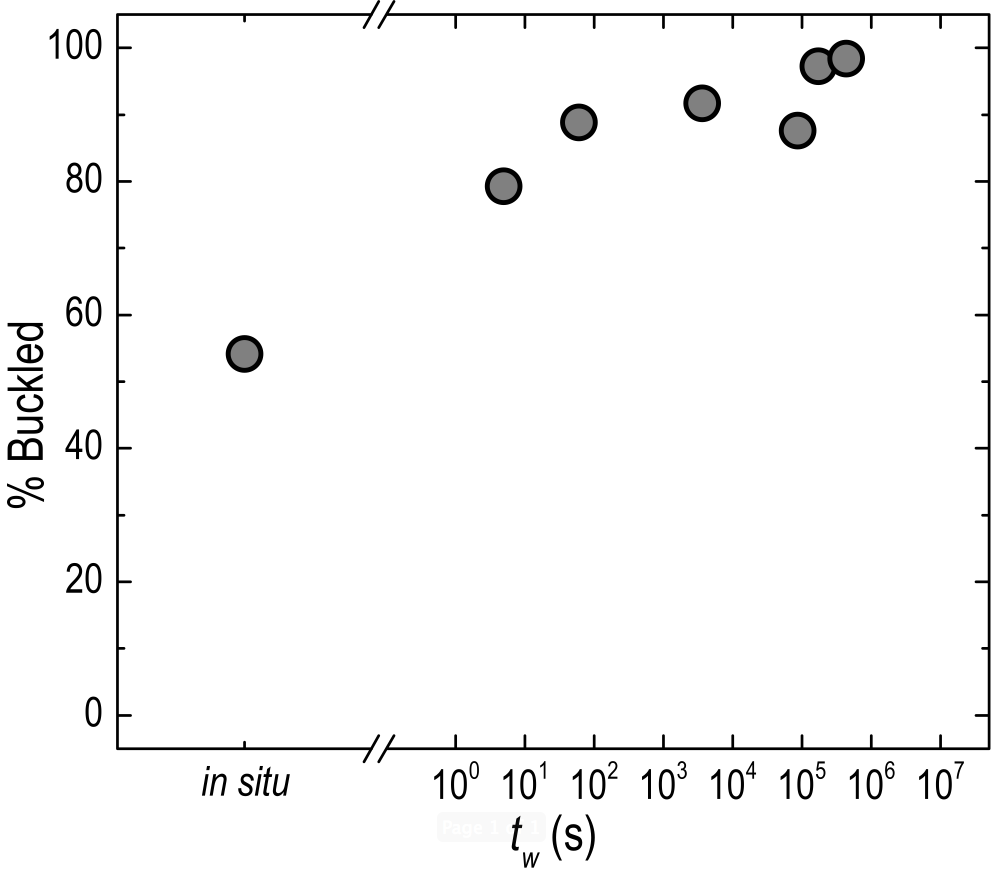}
\caption{Total fraction of capsules that ultimately buckle increases with wait time before polymerization $t_{w}$, and hence the shell inhomogeneity. Capsules have $h_{0}=1.2\mu$m and $R_{0}=70\mu$m, and are buckled at $\Pi\approx0.86$ MPa$>\Pi^{*}$.} 
\end{center}
\end{figure}

\section{Flow rate out of an inhomogeneous capsule}
The buckling of a capsule is driven by the fluid ejection from the capsule interior, due to the imposed osmotic pressure difference across the capsule shell, $\Pi$. This is resisted by the mechanical pressure required to compress the shell, $\Pi_{m}$, at most $\Pi_{m}\approx2E(1-\nu)^{-1}h_{0}(1/R^{*}-1/R_{0})\approx3E(h_{0}/R_{0})[(1-\Delta V^{*}/V_{0})^{-1/3}-1]$, where $R^{*}$ is the radius of the shell at buckling. For the shells studied in this work, $\Pi_{m}<\Pi^{*}$; we thus expect the buckling dynamics to be dominated by the imposed osmotic pressure for $\Pi\gg\Pi^{*}$, and we neglect $\Pi_{m}$ in the simple model presented here. A full treatment of the buckling dynamics that explicitly includes $\Pi_{m}$ confirms the validity of this simplification for the range of $\Pi$ studied here \cite{vella}.

We now estimate the flow rate out of an inhomogeneous capsule due to $\Pi$; the capsule has shell thickness $h(\theta)=h_{0}-\delta\mbox{cos}\theta$, as shown in Fig. 1(a) of the main text. We use spherical coordinates $(r,\theta,\phi)$ centered on the capsule center. For an arbitrary area element $\mbox{d}A$ on the shell surface at $(R_{0},\theta,\phi)$, the local volumetric ejection rate is given by Darcy's law, $\mbox{d}A\cdot\Pi k/\mu h(\theta)$, where $k$ is the shell permeability and $\mu$ is the fluid viscosity. Integrating this over the entire shell surface yields the total ejection rate through the shell:
\begin{multline} \label{eqn0flowrate}
Q=R_{0}^{2}\int_0^{2\pi}\!\mbox{d}\phi\int_0^\pi\!\mbox{sin}\theta\frac{\Pi k}{\mu(h_{0}-\delta\mbox{cos}\theta)}\mbox{d}\theta=\\
\frac{4\pi R_{0}^{2}\Pi k}{\mu h_{0}}\cdot\frac{1}{2\delta/h_{0}}\mbox{ln}\left(\frac{1+\delta/h_{0}}{1-\delta/h_{0}}\right)
\end{multline}

The time delay before the onset of buckling, $\tau$, is the time taken for the volume of fluid ejected from the shell to equate to the threshold buckling volume,  $\tau = \Delta V^{*}/Q$. From Eq.~\ref{eqn0vcdelta} and Eq.~\ref{eqn0flowrate}, we obtain
\begin{equation} \label{eqn-tau}
\tau\approx\frac{V_{0}}{Q_{0}}\sqrt{\frac{3(1-\nu)}{1+\nu}}\frac{h_{0}}{R_{0}}\left(1-\frac{\delta}{h_{0}}\right)^{2}
\end{equation}
where $V_{0} = 4\pi R_{0}^{3}/3$ is the initial capsule volume, $Q_{0}\equiv4\pi R_{0}^{2}\Pi k/\mu h_{0}$ and $k$ is the shell permeability. This is Eq. (2) in the main text.

\section{Estimate of capsule permeability}
\begin{figure}[h]
\begin{center}
\includegraphics[width=3.35in]{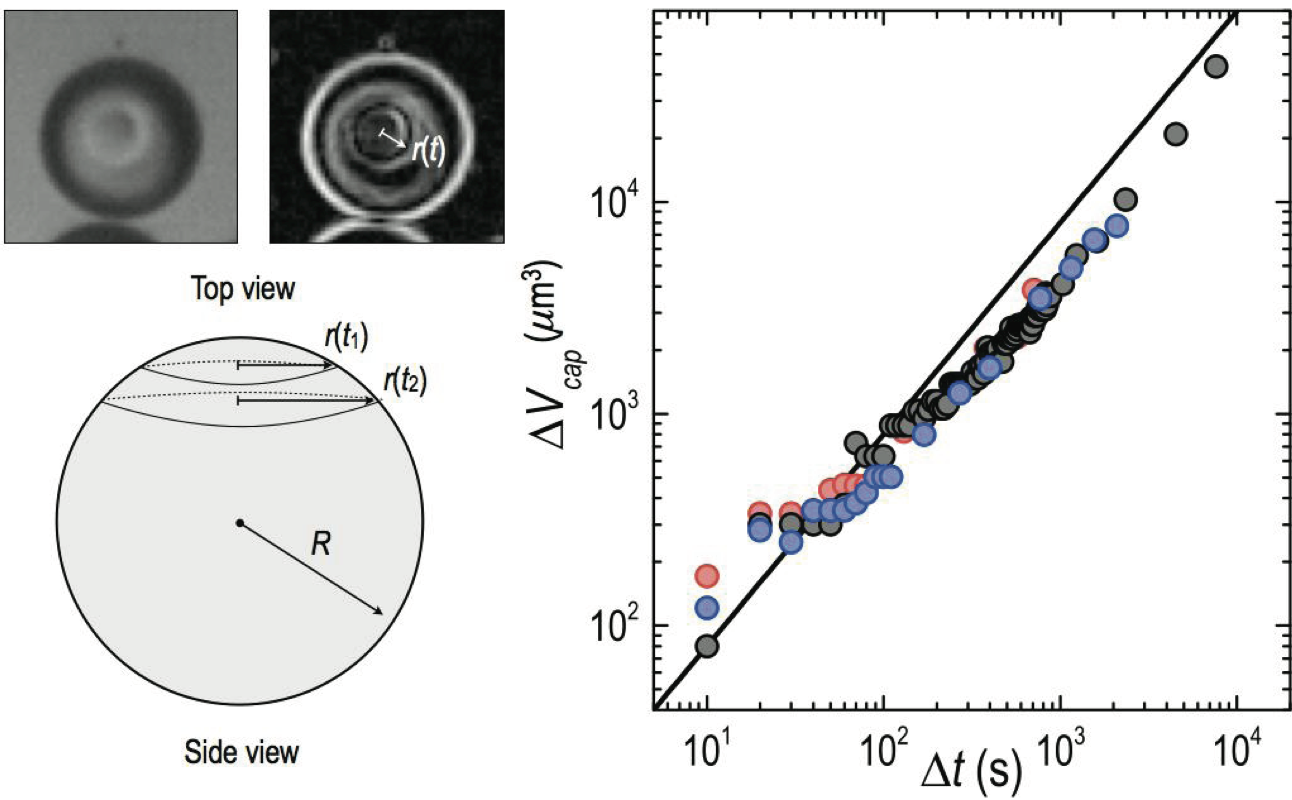}
\caption{Change in the volume of a circular indentation formed in a capsule, $\Delta V_{cap}$, over time $\Delta t$. The indentation forms at $\Delta t=0$. Top left panel shows the top view of the indentation formation, measured using optical microscopy; we use image processing to detect the edge of the indentation, shown in the panel to the right, and track the radius of the indentation over time, $r(t)$. We assume a spherical cap geometry, schematically shown in the side view, to calculate $\Delta V_{cap}$; the data are shown for three different capsules (different colors). We fit the small-time dynamics ($\Delta t<10^{2}$s) to measure the permeability. The capsules have $h_{0}/R_{0}=0.1$ and $\delta/h_{0}\approx0.2$, and are buckled at $\Pi=10$ MPa.} \label{fig-permeability}
\end{center}
\end{figure}

We estimate the capsule permeability $k$ by measuring the change in the radius of a circular indentation $r$ over time, immediately after it is formed in the shell, as shown in Fig.~\ref{fig-permeability}. We assume the volume of the indentation $\Delta V_{cap}$ is given by that of a spherical cap having radius $r(t)$. The permeability is then given by $k\approx\frac{\mu h_{0}d(\Delta V_{cap})/d(\Delta t)}{\Pi\cdot4\pi R_{0}^2}$. In reality, the edges of the indentation are rounded, with radius of curvature $\sim\sqrt{h_{0}R_{0}}\sim15\mu$m \cite{landau}; we thus expect our estimated $\Delta V_{cap}$ to overpredict the actual $\Delta V$, and consequently, we expect to underpredict the shell permeability by a factor $\sim2$.

\section{Lubrication theory of the formation of the inhomogeneous shell (Communicated By Howard A. Stone, February 2012)}
To estimate the variation of the shell inhomogeneity, $\delta/h_{0}$, with time, $t_{w}$, we apply lubrication theory to the double emulsion geometry shown in Fig. 1(a) of the main text. Before UV polymerization at time $t_{w}$, the inner droplet containing 10wt\% PVA has radius $R_{0}-h_{0}$ and the outer droplet containing ETPTA monomer has radius $R_{0}$; the shell thickness is then $h(t_{w})=h_{0}-\delta(t_{w})\mbox{cos}\theta$. The droplets are collected in 8 wt\% or 10 wt\% PVA solution. We note that, for our shells, $h_{0}/R_{0}\leq0.1$. Moreover, using our experimental measurements of $\delta(t_{w})$, we estimate the characteristic translation speed of the inner droplet as $\sim1\mu$m/day; using a characteristic droplet length scale $<100\mu$m, shell viscosity $\mu_{o}=65$mPa.s, fluid-fluid surface tension $\sim2.5$mN/m, this corresponds to a Reynolds number $Re<10^{-10}$ and a Capillary number $Ca<10^{-10}$. The fluid-fluid interface can thus be approximated as spherical, and the flow in the shell between the inner and outer droplets is well described using lubrication theory. We denote $x=r\theta$ as the direction along the shell, where $r$ is the spherical radial distance measured from the center of the inner sphere, and $z$ is across the shell. 

Because both inner and outer droplets are stabilized by 8-10 wt\% PVA, Marangoni stresses resist shear stresses at the fluid interfaces; thus, we assume no-slip boundary conditions on the fluid-fluid interfaces at $z=0$ and $z=h(\theta,t_{w})$. The Stokes equation then yields the velocity distribution: 
\begin{equation}
u_{x}(\theta,z,t_{w})=\frac{1}{2\mu(R_{0}-h_{0})}\frac{\partial p}{\partial\theta}z(z-h(\theta,t_{w}))
\end{equation}

Integrating the continuity equation twice across the gap, with $u_{r}=0$ at the outer boundary, $u_{r}=U\mbox{cos}\theta$ at the inner translating boundary, and $u_{\theta}=u_{x}=0$ at both boundaries, we find the pressure distribution
\begin{equation}
p(\theta,t_{w})=p_{0}(t_{w})-\frac{3\mu_{o}(R_{0}-h_{0})^{2}U}{\delta h^{2}}
\end{equation}

Balancing forces in the $z$ direction, $2\pi\int_{\theta=0}^{\pi}p{\bf n\cdot e_{z}}~\mbox{sin}\theta\mbox{d}\theta=F^{b}$, where $F^{b}\equiv(4/3)\pi(R_{0}-h_{0})^{3}\Delta\rho g$ is the buoyant force on the inner sphere and $\Delta\rho$ is the difference in density between the inner and outer spheres (0.08 g/cm$^{3}$). We complete the integral and substitute $U=\partial\delta/\partial t_{w}$, and take the limit $\delta/h_{0}\rightarrow1$ \cite{howard}; we obtain
\begin{multline}
\frac{1}{h_{0}-\delta}\frac{\partial\delta}{\partial t_{w}}=\frac{F^{b}h_{0}^{2}}{6\pi\mu_{o}(R_{0}-h_{0})^{4}}\\\implies\frac{\delta(t_{w})}{h_{0}}=1-e^{-\frac{2\Delta\rho gh_{0}^{2}t_{w}}{9\mu_{o}(R_{0}-h_{0})}}
\end{multline}

\begin{figure}[h]
\begin{center}
\includegraphics[width=3.5in]{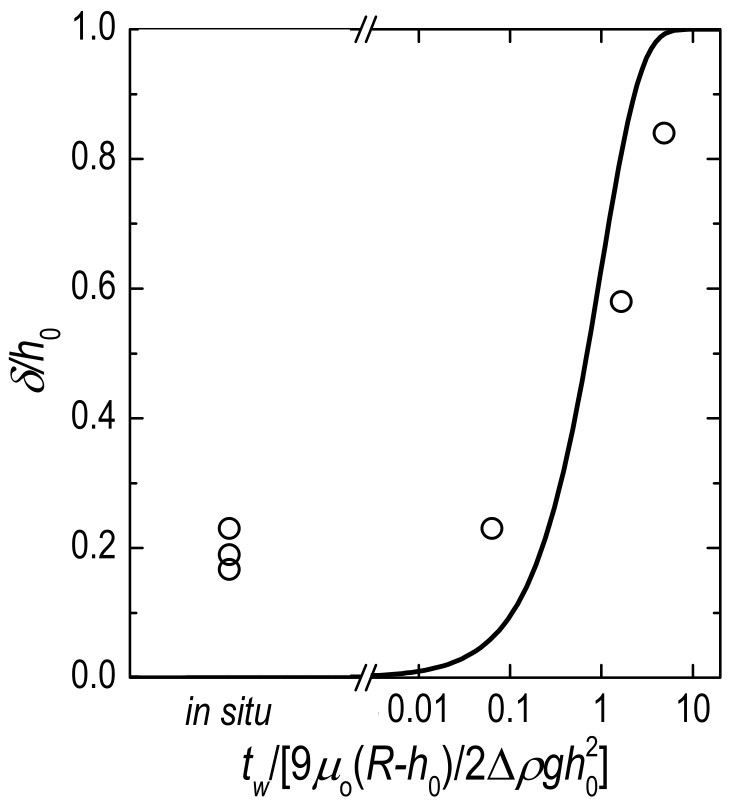}
\caption{Measurements of $\delta/h_{0}$ as a function of $t_{w}\cdot\frac{2\Delta\rho gh_{0}^{2}}{9\mu_{o}(R-h_{0})}$, along with lubrication theory solution (black line), showing agreement between the two.} \label{fig-lubrication}
\end{center}
\end{figure}

To test this prediction, we use SEM or confocal microscopy measurements of capsules with fluorescently-labeled shells to directly measure $\delta/h_{0}$ for capsules of varying $t_{w}$. We find good agreement between the two, particularly for large $t_{w}$, when $\delta/h_{0}$ precipitously rises to 1, as shown in Fig.~\ref{fig-lubrication}. We find $\delta/h_{0}\rightarrow0.2$ as $t_{w}\rightarrow0$, in contrast to the theory; this is to be expected, due to the slight bias in the inner droplet position as the double emulsions are formed in the microfluidics device.

\section{Numerical simulations of thin inhomogeneous shells} \label{sec-simulations}
To test our theoretical predictions and to understand the postbuckling shapes of the capsules, we perform numerical simulations of a variety of thin, elastic, spherical shells whose surfaces are composed of ``amorphous" randomly-positioned points connected by springs. Similar models have been used to study the deformations of uniform spherical shells \cite{kato,gompper}, viruses \cite{lidmar,gompper2,lenz,lenz2,siber} and pollen grains \cite{eleni}.

We simulate shells with a non-uniform thickness profile, described by $h(\theta)=h_{0}-\delta\mbox{cos}\theta$, similar to the experimental capsules; $\theta$ is the inclination angle from the thinnest point of the shell. The spring stiffnesses are chosen so that the 3D Young's modulus of the shell is $E=600$ MPa and the Poisson ratio is $\nu=1/3$, similar to the material making up the polymeric capsules. Since $h$ is always much smaller than the radius of the shell, the elastic energy may be described by a sum of bending and stretching energies of an initially spherical two-dimensional elastic sheet \cite{landau,niordson}:
\begin{multline}
E=\\
\int dS[\frac{E h(\theta)}{2(1-\nu^{2})}\cdot\left((u_{11}+u_{22})^{2}-2(1-\nu)(u_{11}u_{22}-u_{12}^{2})\right)\\
+\frac{Eh(\theta)^{3}}{24(1-\nu^{2})}\cdot\left((k_{11}+k_{22})^{2}-2(1-\nu)(k_{11}k_{22}-k_{12}^{2})\right]
\end{multline}
where $u_{ij}$ is the $2\times2$ strain tensor, $k_{ij}$ is the change in the curvature tensor from its initial value, and the integration is carried out over the mid-surface of the shell. The initial curvature tensor $k^{0}_{ij}$ of the shell mid-surface is that of a sphere with radius $r$: $k^{0}_{11}=k^{0}_{22}=1/r;k^{0}_{12}=0$ at every point for an orthonormal basis set up in the tangent plane to the sphere at that point.

To numerically simulate the shells, we discretize the elastic energies on a mesh of 20,000 points. The initial, unstrained configuration is obtained by distributing the points quasi-randomly on the surface of a sphere, maintaining a minimum distance between nearest neighbors. The disorder in the initial mesh eliminates the effect of the 12 regularly spaced five-fold disclinations that inevitably arise when attempting to cover a spherical surface with equilateral triangles \cite{lidmar}. Bonds are drawn between nearest neighbor pairs to form a mesh; the topology of the mesh is unchanged after initialization (no rearrangement of bonds takes place), consistent with the polymerization imposed in the experiments. A point is arbitrarily chosen to be the thinnest point of the sphere. The elastic stretching energy of deformations from the initial unstrained configuration is approximated by a harmonic spring energy associated with each bond \cite{seung}: 
\begin{equation}
E_{s}=\sum_{\langle ij\rangle}\frac{\sqrt{3}}{4}Eh(\theta_{ij})(r_{ij}-r_{ij}^{0})^{2}
\end{equation}
where $r_{ij}$ and $r_{ij}^{0}$ are the lengths in the deformed and initial states of the bond connecting nearest-neighbor mesh points $i$ and $j$, and $\theta_{ij}$ is the inclination angle between the centre of the spring and the thinnest point. The spring constant is chosen to reproduce the thickness-dependent elastic modulus in the continuum limit, with $\nu=1/3$ \cite{seung}.

The often-used discretization of the bending energy in terms of the angles between normals of adjacent facets in the mesh is not suitable for disordered meshes where all facets are not equilateral triangles \cite{kroll}. Furthermore, this discretization scheme does not allow for a nonzero background curvature. Here we use a different approach. The curvature tensor is constructed from the mean curvature $H$ and Gaussian curvature $K$ (respectively, the trace and the determinant of the curvature tensor) associated with each point in the deformed state. Note that the Gaussian curvature energy is {\it not} an invariant for inhomogeneous shells, and thus cannot be ignored, in contrast to uniform shells \cite{julicher}. An approximation for the mean curvature associated with site $i$ is \cite{kroll}:
\begin{equation}
H_{i}=\frac{1}{\sigma_{i}}\left|\sum_{j(i)}\frac{\sigma_{ij}}{r_{ij}}({\bf R}_{i}-{\bf R}_{j})\right|
\end{equation}
where $\sigma_{ij}=r_{ij}(\mbox{cot}\theta_{1}+\mbox{cot}\theta_{2})/2$ is the length of a bond in the {\it dual mesh} connecting the centers of the two triangular facets $ijk$ and $ijk'$ that share the bond linking points $i$ and $j$ (calculated in terms of the interior angles $\theta_{1}, \theta_{2}$ of the vertices $k$ and $k'$), $\sigma_{i}=\sum_{j(i)}\sigma_{ij}r_{ij}/4$ is the area of the dual cell (the polygon of bonds of the dual mesh surrounding vertex $i$), ${\bf R}_{i}$ is the three-dimensional coordinate vector of point $i$, and all sums are over the neighbors of site $i$. The Gaussian curvature is approximated by exploiting the Gauss-Bonnet theorem \cite{meyer}:
\begin{equation}
K_{i}=\frac{1}{\sigma_{i}}\left(2\pi-\sum_{j(i)}\alpha_{j}\right)
\end{equation}
where $\alpha_{j}$ is the angle facet $j$ subtends at point $i$ and the sum runs over all facets sharing point $i$. The discretized version of the bending contribution to the elastic energy [Eq. S3] is
\begin{multline}
E_{b}=\\
\sum_{i}\frac{Eh(\theta)^{3}\sigma_{i}}{24(1-\nu^{2})}\left[\left(H_{i}-\frac{2}{r}\right)^{2}-2(1-\nu)\left(K_{i}-\frac{H_{i}}{r}+\frac{1}{r^{2}}\right)\right]
\end{multline}

The total energy $E_{s}+E_{b}$, a function of the $3\times20,000$ variables describing the point positions, is numerically minimized using the BFGS quasi-Newton optimization algorithm implemented in the freely available GNU Scientific Library \cite{galassi}. 

To numerically obtain the threshold buckling pressure from simulations, we add a term $E_{p} = \Pi V$ to the discretized elastic energy energy, where $V$ is the volume enclosed by all the facets of the mesh. The pressure $\Pi$ is gradually increased until the shell collapses, and the corresponding threshold pressure $\Pi^{*}$ identified. 

To find energy-minimizing configurations at a particular target volume $V_{T}$, we use the penalty method \cite{siber2,eleni}. We add a penalizing term $E_{vol}=\lambda(V-V_{T})^{2}$ to the energy, where $V$ is the volume enclosed by all the facets of the mesh, with $\lambda$ initially chosen to be very small. We increment $\lambda$ systematically between successive minimizations of the total energy, until the difference between the actual and target volumes is negligible. The final configuration is checked for stability to a small random perturbation of all the point positions.

To find the threshold volume reduction at collapse, we initially set $V_{T}$ to the volume $V_{0}$ enclosed by a sphere of radius $r$. We then reduce the target volume by small amounts, finding the minimum energy configuration after each decrement. The volume decrement between steps is $0.00025V_{0}$.
To quantify the buckling, we define the asphericity of the shell as \cite{lidmar}
\begin{equation}
\frac{\langle\Delta R^{2}\rangle}{\langle R\rangle^{2}}=\frac{1}{N}\sum_{i}\frac{(R_{i}-\langle R\rangle)^{2}}{\langle R\rangle^{2}}
\end{equation}
where $R_{i}$ is the radial distance of point $i$ and $\langle...\rangle$ denotes averaging over all points of the mesh. The discrete nature of the buckling inversion can be captured by plotting the asphericity of the final shape against $\Delta V/V_{0}$: the asphericity first becomes nonzero at a finite volume reduction which we identify as $\Delta V^{*}$.

\subsection{Postbuckling shapes of shells}
To investigate the influence of inhomogeneity on the shape of the inversion at large volume reductions, we simulate large volume reductions for two sets of geometric parameters: shells with $h_{0}/R=0.018$ and $\delta/h_{0}=0.20$, similar to the more homogeneous shells described in the main text, and shells with $h_{0}/R=0.0165$ and $\delta/h_{0}=0.82$, similar to the very inhomogeneous shells. We use the penalty method outlined in the previous subsection to reduce the enclosed volume in steps of $0.0025V_{0}$ from $V_{T}=V_{0}$ to $V_{T}=0.75V_{0}$. For each set of geometric parameters, we investigate the robustness of the resulting shapes by performing eight different simulations, in each case varying the point on the underlying mesh that corresponds to the thinnest point of the shell. This effectively changes the random discretization of the shell in each instance.

In all instances, the shells buckle when the shell volume is reduced past $\Delta V^{*}$ (whose value is approximately $0.0115 V_{0}$ for the more homogeneous shells and $0.0008V_{0}$ for the very inhomogeneous shells) to form a single circular indentation centered at the thinnest part of the shell. At larger volume changes, however, the shape of the inversion differs between the two shells and among different simulation instances for the same shell. In more homogeneous shells, the inversion remains circular upto a volume change of $\Delta V \approx 0.05 V_{0}$, beyond which the perimeter sharpens into straight ridges connected by 4-5 vertices [main text, Fig. 3(d)]. Three out of the eight simulation runs on the more homogeneous shells displayed inversions with 4 vertices, while the remaining five runs displayed inversions with 5 vertices. Remarkably, the two types of inversions do not differ significantly in their elastic energies, which explains why slight differences in the underlying mesh have a significant effect on the final shape.

The shape evolution of the inversions in very inhomogeneous shells is qualitatively different [main text, Fig. 3(e)]. The inversion takes on a ridged appearance almost immediately after its formation, at a much smaller volume change ($\Delta V \approx 0.004 V_{0}$) compared to the more homogeneous shells. The number of vertices in the perimeter is also consistently higher for the very inhomogeneous shells: out of eight simulation runs, three displayed inversions with 5 vertices, four had inversions with 6 vertices, and one run displayed an inversion with 7 vertices. Again, there is no significant difference in the elastic energies of the competing shapes.

We note that while the exact number of vertices formed differs between the experiments and the numerical simulations, both show that the inversions in very inhomogeneous capsules develop more vertices than the more homogeneous case.

We further characterize the dependence of postbuckling shapes on shell inhomogeneity by simulating large volume reductions for shells with $h_0/R=0.2$, similar to the experimental capsules, but varying $\delta/h_0$ from 0.05 to 0.95. The number of vertices observed in the inversion is shown in Fig. S10. We find that the number of vertices increases with inhomogeneity and with increasing volume reduction.

\begin{figure}[h]
\begin{center}
\includegraphics[width=3.15in]{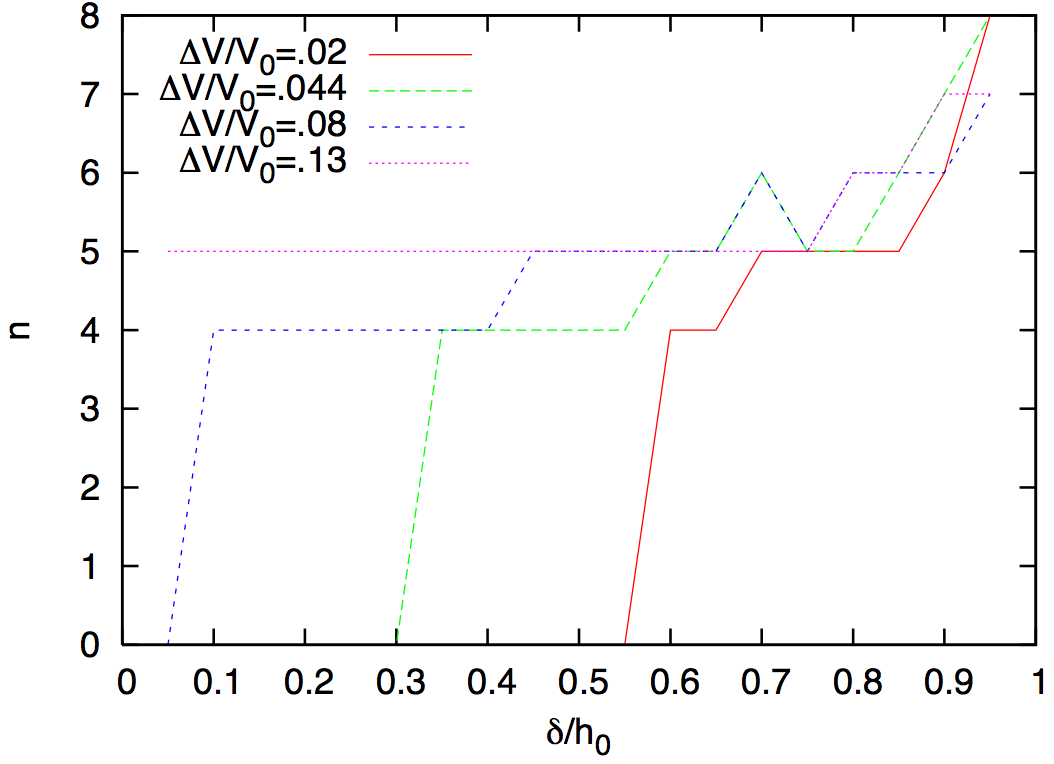}
\caption{Inhomogeneity dependence of the postbuckling shape. The number of vertices, $n$, of the polygonal inversion is shown for four volume reductions as a function of the shell inhomogeneity $\delta/h_0$. A value of $n=0$ corresponds to a roughly circular inversion with no discernible sharp points.
 For all shells, $h_0/R = 0.02$. The volume was reduced in steps of $0.005V_0$ from $V_T=V_0$ to $V_T=0.87V_0$. We used the same mesh for all the simulations.} 
\end{center}
\end{figure}

\section{Micrographs of buckled capsules}
\begin{figure}[h]
\begin{center}
\includegraphics[width=3.15in]{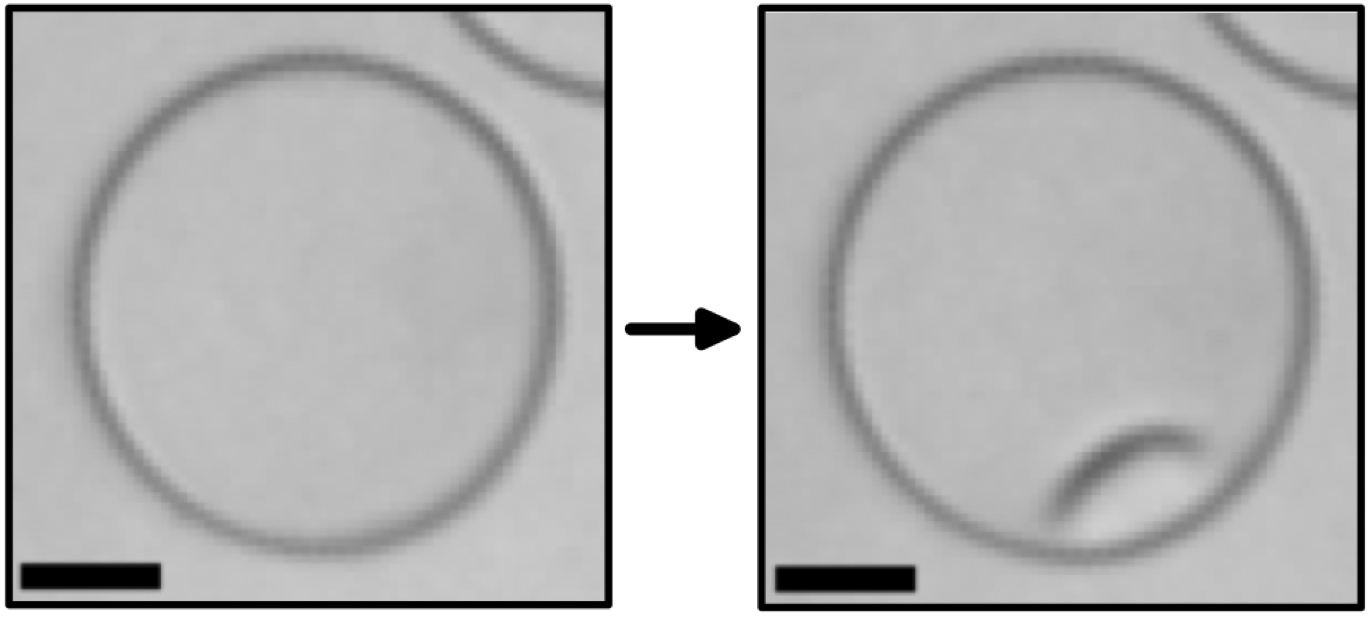}
\caption{Optical micrograph of the buckling of a capsule; scale bar is $20\mu$m. } 
\end{center}
\end{figure}

\begin{figure}[h]
\begin{center}
\includegraphics[width=3.15in]{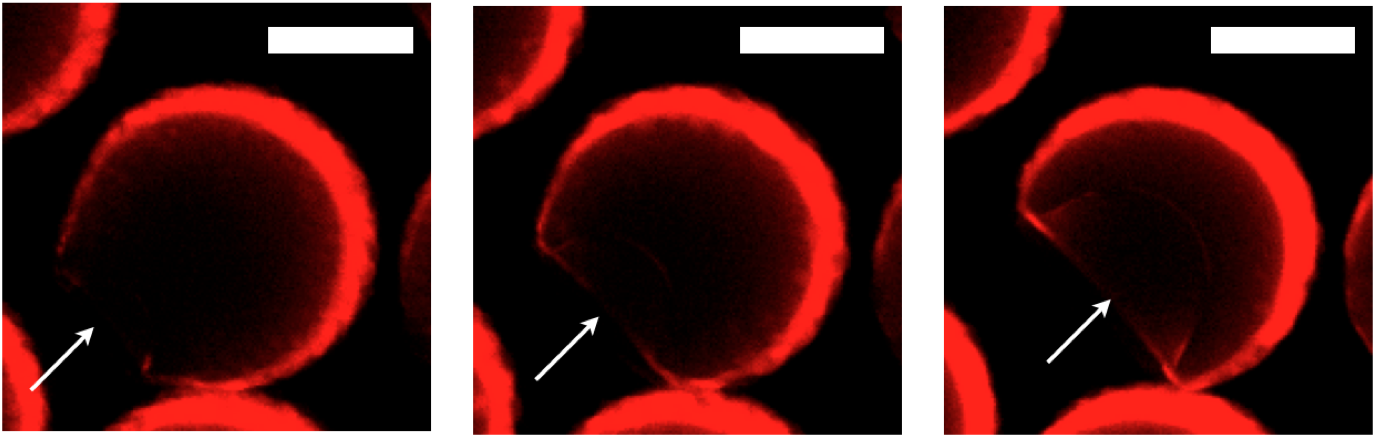}
\caption{Buckling begins at the thinnest part of the shell for capsules with thickness inhomogeneity $\delta/h_{0}\approx0.84$}; scale bars are $50\mu$m. These capsules are prepared with fluorescent shells to enable visualization with confocal microscopy.
\end{center}
\end{figure}

\begin{figure*}[h]
\begin{center}
\includegraphics[width=7.15in]{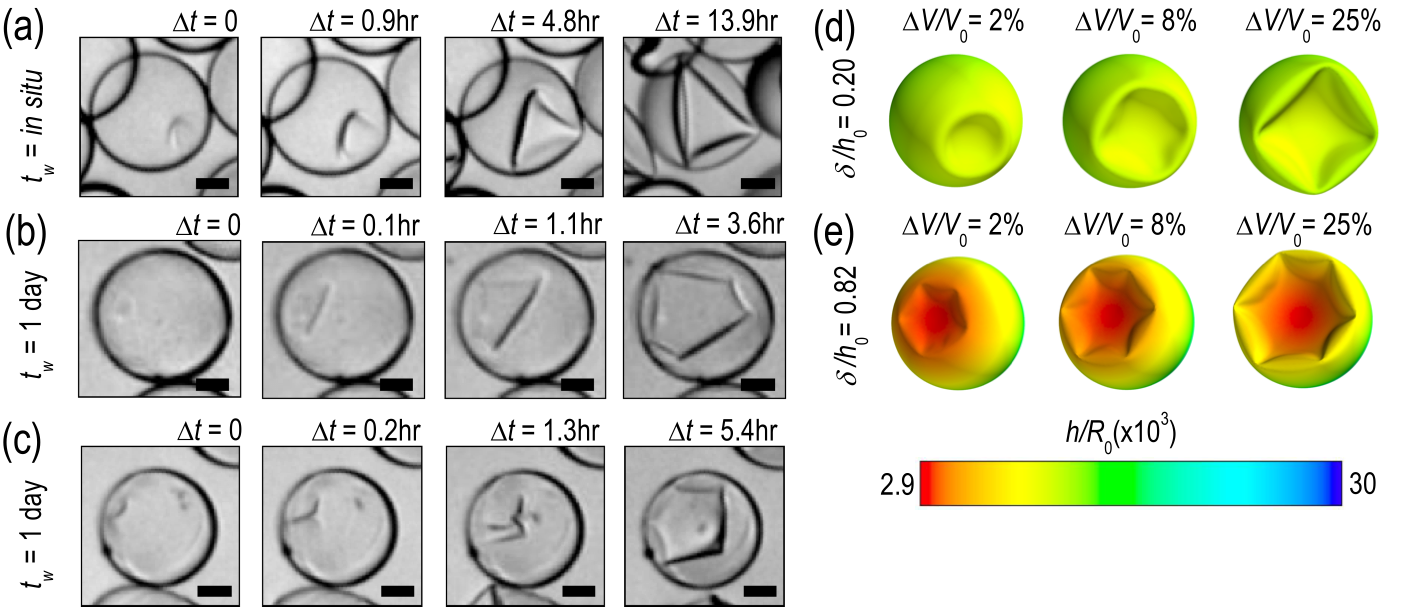}
\caption{Folding pathways for different shell inhomogeneities. (a-c) Optical microscope images exemplifying buckling at $\Pi\approx0.86$ MPa of (a) slightly inhomogeneous capsules polymerized {\it in situ} ($t_{w}\approx0$), with $\delta/h_{0}\approx0.2$, (b-c) very inhomogeneous capsules polymerized after a wait time $t_{w}=1$ day, with $\delta/h_{0}\approx0.84$. Very inhomogeneous capsules buckle through the formation of either (b) one single indentation or (c) two indentations. $\Delta t$ is time elapsed after buckling. Scale bars are $35\mu$m. (d-e) Examples of simulated shells with similar geometries as the capsules shown in (a-c), for varying fractional volume reduction $\Delta V/V_{0}$. Color scale indicates the spatially-varying shell thickness.} 
\end{center}
\end{figure*}